\documentclass[12pt,preprint]{aastex}
\usepackage{epsf,color,cancel}

\newcommand{\rhozero}{\rho_0}

\newcommand{\gammabar}{\overline{\Gamma}_1}

\newcommand{\rhocutoff}{\rho_\mathrm{cutoff}}

\newcommand{\gcc}{\mathrm{g~cm^{-3} }}
\newcommand{\cms}{\mathrm{cm~s^{-1} }}

\newcommand{\nablab}{\mathbf{\nabla}}
\newcommand{\cdotb}{\mathbf{\cdot}}

\newcommand{\sfrac}[2]{\mathchoice
  {\kern0em\raise.5ex\hbox{\the\scriptfont0 #1}\kern-.15em/
   \kern-.15em\lower.25ex\hbox{\the\scriptfont0 #2}}
  {\kern0em\raise.5ex\hbox{\the\scriptfont0 #1}\kern-.15em/
   \kern-.15em\lower.25ex\hbox{\the\scriptfont0 #2}}
  {\kern0em\raise.5ex\hbox{\the\scriptscriptfont0 #1}\kern-.2em/
   \kern-.15em\lower.25ex\hbox{\the\scriptscriptfont0 #2}}
  {#1\!/#2}}

\newcommand{\half}{\sfrac{1}{2}}

\newcommand{\myhalf}{\sfrac{1}{2}}

\newcommand{\dt}{\Delta t}

\newcommand{\Hnuc}{H_{\rm nuc}}

\newcommand{\er}{\mathbf{e}_r}
\newcommand{\ex}{\mathbf{e}_x}
\newcommand{\ey}{\mathbf{e}_y}
\newcommand{\ez}{\mathbf{e}_z}

\newcommand{\ubold}{\mathbf{U}}
\newcommand{\ut}{\widetilde{\ubold}}

\newcommand{\x}{\mathbf{x}}

\newcommand{\omegadot}{\dot{\omega}}

\newcommand{\favgradvel}{\langle \tilde{v}_r \rangle}

\begin{document}
\title{Low Mach Number Modeling of Type Ia Supernovae.  IV.  White Dwarf Convection}
\shorttitle{X-Ray Bursts}
\shortauthors{Almgren et al.}

\author{M.~Zingale\altaffilmark{1},
        A.~S.~Almgren\altaffilmark{2},
        J.~B.~Bell\altaffilmark{2},
        A.~Nonaka\altaffilmark{2},
        S.~E.~Woosley\altaffilmark{3}}

\altaffiltext{1}{Dept. of Physics \& Astronomy,
                 Stony Brook University,
		 Stony Brook, NY 11794-3800}

\altaffiltext{2}{Center for Computational Science and Engineering,
                 Lawrence Berkeley National Laboratory,
                 Berkeley, CA 94720}

\altaffiltext{3}{Dept. of Astronomy  \& Astrophysics,
                 The University of California, Santa Cruz,
                 Santa Cruz, CA 95064}

\begin{abstract}
We present the first three-dimensional, full-star simulations of
convection in a white dwarf preceding a Type Ia supernova, specifically
the last few hours before ignition.  For these long-time calculations
we use our low Mach number hydrodynamics code, MAESTRO, which
we have further developed to treat spherical stars centered in a 
three-dimensional Cartesian geometry.  
The main change required is a procedure to map the
one-dimensional radial base state to and from the Cartesian grid.  Our
models recover the dipole structure of the flow seen in previous
calculations, but our long-time integration shows that the orientation
of the dipole changes with time.  Furthermore, we show the development
of gravity waves in the outer, stable portion of the star. 
Finally, we evolve several calculations to the point of ignition
and discuss the range of ignition radii.
\end{abstract}
\section{Introduction}
Modeling highly subsonic convection in stars requires algorithms
designed for long time integration.  In the low Mach number
approximation, we filter out sound waves while keeping the
compressibility effects important to describing the flow.  In our
previous work (see \citealt{ABRZ:I}---henceforth paper I,
\citealt{ABRZ:II}---henceforth paper II, and
\citealt{ABNZ:III}---henceforth paper III), we developed a low Mach
number stellar hydrodynamics algorithm for reacting full-star flows in
order to study the convective phase of Type Ia supernovae (SNe Ia).
In paper I, we derived the low Mach number equation set.  In paper II,
we included the effects of heat release due to external sources and
allowed for a time-dependent background state.  In paper III, we
incorporated reactions into the system and also allowed the background
state to evolve in response to large-scale convection and large-scale
heating.  Here, we extend the algorithm to spherical full-star problems 
using a three-dimensional Cartesian grid geometry.


Our target application for this algorithm is the period of convection
that precedes the ignition of SNe Ia.  The standard model of a SN Ia 
involves a white dwarf in a binary system accreting from a normal
companion, and approaching the Chandrasekhar mass (see for example
\citealt{hillebrandtniemeyer2000}).  The increase in the central
temperature and density accompanying the accretion seed carbon burning
in the core, which in turn drives convection in the star.  This
convective `simmering' phase can last centuries, slowly increasing the
core temperature of the white dwarf
\citep{Woosley:2004,wunschwoosley:2004}.  
A similar starting condition might be achieved in merging white dwarfs
if mass is added slowly enough to avoid ignition at the edge of the stars
\citep{yoon2007}. 
During this phase, fluid
heated by reactions buoyantly rises and cools via expansion,
exchanging heat with its surroundings.  The extent of the convective
region grows with increasing temperature, eventually covering
roughly the inner solar mass of the star.  Outside of the convective
region, the star is stably stratified.

The continued increase in central temperature, coupled with the extreme
temperature sensitivity of the carbon reactions, means that eventually
the reactions proceed vigorously enough that a hot bubble cannot
cool fast enough, and a burning front is born.  This happens for a
temperature of about $7-8\times 10^8$~K \citep{nomoto1984iii,woosley1990}.  This 
burning front will quickly propagate through the white dwarf,
converting most of the carbon/oxygen fuel to heavier elements, and
releasing enough energy to unbind the star.  However, exactly where in
the star the ignition takes place is still unknown.  Among the
earliest work to consider the role of buoyancy in off-center ignition
were \citet{garciasenzwoosley1995}, \citet{bychkovliberman}, and \citet{niemeyer:1996}. 
Additionally, some multidimensional studies have been done of the dynamics of the first
bubbles to ignite in a white dwarf \citep{iapichino,zingaledursi}.
These papers made the case that we really need to understand whether
the ignition is at the center or off-center.  As calculations have
become more sophisticated, it has only become more clear that 
the outcome of the explosion is extremely sensitive to exactly how the
burning fronts are initiated
\citep{gamezo:2005,jordan:2008,roepke2007,garciasenz:2005}.

Less work has been done on multidimensional modeling of the convective
phase preceding the explosion.  To date, no multidimensional
calculation of the convection in the white dwarf has modeled the
entire star.  The major contributions thus far are two-dimensional
simulations of a 90$^\circ$ wedge of the star using an implicit
hydrodynamics code \citep{hoflichstein:2002,steinwheeler2006}, and a
three-dimensional anelastic calculation \citep{kuhlen-ignition:2005}
of the inner convective region of the star.  All of these calculations
cut out a small part of the central region of the star to avoid the
coordinate singularity at the origin in spherical coordinates.
Furthermore, the Kuhlen et al.\ calculation modeled the star out to a
radius of only 500~km---leaving out part of the convective zone and
the surrounding, stably stratified region.  The calculations by
\citet{hoflichstein:2002} found ignition near the center of the white
dwarf, produced by the fluid flow converging toward the center,
with convective velocities of about
100~km~s$^{-1}$.  However, the ignition they see was likely affected
by the converging geometry of their computational
domain.  The three-dimensional calculations by
\citet{kuhlen-ignition:2005} showed that the large-scale flow took on
a dipole pattern, suggesting that off-center ignition in an {\em
  outflow} on one side of the star might be
favored.  They also investigated the role
of rotation.  Finally, recent calculations shown in
\cite{woosley-scidac2007} used an anelastic method on a Cartesian
grid, avoiding the singularity in the center, but still cut out the
outer part of the convective region and the convectively stable region
surrounding it.  Here the dipole was once again seen.

As seen from the wide range of explosion outcomes in the literature,
realistic initial conditions are a critical part of SNe Ia
modeling.  Only simulations of this convective phase can yield the
number, size, and distribution of the initial hot spots that seed the
flame.  Additionally, the initial turbulent velocities in the star are
at least as large as the laminar flame speed \citep{hoflichstein:2002}, 
so accurately representing this initial flow may be an important
component to explosion models.  Perhaps owing to a limited number of
convection calculations, with few exceptions \citep{livne2005}, nearly
all explosion models to date begin with a quiet (zero velocity) white
dwarf.

Our goal in this study is to demonstrate that we have developed low
Mach number hydrodynamics to the point where we can perform detailed
calculations of the convective flow preceding the explosion, and to
begin to understand the nature of the dynamics.  In this work, we
model the entire star, including the region surrounding the convective
zone.  Recently, it has been suggested \citep{pirochang} that the
dynamics at the interface between the convective and stably-stratified
regions of the star may be important during the flame propagation
phase.  Only full star calculations can capture this part of the flow.
The resulting simulations can then form the basis for simulations of
the flame propagation to build a more detailed picture of SNe Ia.

\section{Numerical Methodology and Setup}

The basic idea of low Mach number hydrodynamics is to reformulate
the fluid equations to filter out sound waves while retaining the
compressibility effects important to the problem---in this case, local
compressibility effects due to burning, and large scale effects due to
the background stratification of the star.  A full derivation of the
equations of low Mach number hydrodynamics is presented in papers
I--III.  Here we show the final equations and discuss adjustments
needed for the spherical star.  We recall that the use of low Mach
number equations rather than the fully compressible equations enables
the use of a time step based on the fluid velocity rather than
the sound speed;  this allows a $1 / M$ increase in the time step
over traditional compressible codes, where the Mach number, $M,$
represents the ratio of fluid velocity to sound speed.
During the convective phase preceding the first flames in SNe Ia,
we expect the Mach number to be $O(0.01)$, making 
a low Mach number algorithm an appropriate choice.

We choose to discretize our three-dimensional grid using Cartesian
rather than spherical coordinates in order to avoid a coordinate
singularity at the center of the star.  This gives rise to the most
notable difference from paper~III---the base state is a 1-d radial
profile and is not aligned with the any of the axes in the 
three-dimensional Cartesian grid.  We refer
to this as a spherical geometry, reflecting the fact that the base
state is discretized in 1-d spherical coordinates.  Throughout this
paper, we refer to the Cartesian coordinates of the center of the
star as $(x_c,y_c,z_c)$.

\subsection{\label{sec:equationset} Equation Set}

The formulation of our equations relies on the existence of a base
state density, $\rho_0(r)$, and pressure, $p_0(r)$, that are in hydrostatic
equilibrium, $\nablab p_0 = \rho_0 g \er,$ where $\er$ is a unit
vector pointing in the radial direction from the center of the star.  
In spherical geometries, the gravitation acceleration, $g(r)$, 
is computed solely using the base state density as
\begin{equation}
g(r) = -\frac{G M_\mathrm{encl}(r)}{r^2} 
\end{equation}
with the mass enclosed within a radius $r$ defined as
\begin{equation}
M_\mathrm{encl}(r) = 4\pi \int_0^r \rho_0(r') r'^2 dr' \enskip .
\end{equation}
As we discuss in \S~\ref{sec:initial_model}, we use a cutoff density,
$\rhocutoff$, in our initial model.  The star is mapped onto the grid
down to this cutoff density, surrounded by an ambient medium.  In
computing $M_\mathrm{encl}$, we stop contributing to $M_\mathrm{encl}$ once the density drops
below $\rhocutoff$.

In this paper, we reuse much of the notation from paper~III.  An
overbar represents the average of a quantity over a layer of constant
radius in the star, 
\begin{equation}
\overline{\phi}(r) = \frac{1}{\mathrm{A}(\Omega_H)}\int_{\Omega_H} 
       \phi(\x) \; dA \enskip ,
\end{equation}
where $\Omega_H$ is a region at constant radius in the star, and
${\mathrm A}(\Omega_H) \equiv \int_{\Omega_H} dA$.  
In this notation, $\x$ represents the Cartesian coordinates on the 3-d
grid and $r$ is the base state radial coordinate centered at
$(x_c,y_c,z_c)$.  A subscript `0'
represents a base state quantity.  We compute $\er$ in a
cell indexed by $(i,j,k)$ with Cartesian coordinates $(x_i,y_j,z_k)$ as 
\begin{equation}
\er = \frac{x_i - x_c}{r} \ex + \frac{y_j - y_c}{r} \ey + \frac{z_k - z_c}{r} \ez \enskip ,
\end{equation}
with $r^2 = (x_i - x_c)^2 + (y_j - y_c)^2 + (z_k - z_c)^2$, and $\ex$,
$\ey$, $\ez$ the unit vectors for the Cartesian coordinate system.

In our previous work, the total fluid velocity, $\ubold$, was
decomposed into a local velocity field, $\ut$ and base state velocity,
$w_0$, as
\begin{equation}
\ubold = \ut(\x,t) + w_0(r,t) \er \enskip .
\end{equation}
The base state velocity,
is used to adjust the base state in response to the heating on the 
grid.  In paper II, we demonstrated that when the heating is large, 
expanding the base state is critical to accurately modeling the flow.

In the current application, convection in the white dwarf, the heating
is small until the flame ignites.  Therefore,
for these first calculations, we use a background state that is
fixed in time.  We will later quantify the extent to which this assumption
of a fixed background state is valid.  
This simplifies the evolution equations, and we can now
use $\ubold$ for $\ut$ and $w_0 = 0$.

The full state evolves according to 
\begin{eqnarray}
\frac{\partial (\rho X_k)}{\partial t} &=& 
       - \nablab \cdotb (\ubold \rho X_k) +
         \rho {\omegadot}_k \enskip , \label{eq:species} \\
\frac{\partial\ubold}{\partial t} &=& - 
 \ubold \cdotb \nablab \ubold 
      - \frac{1}{\rho} \nablab\pi
      - \frac{(\rho-\rhozero)}{\rho} \; g \; \er  \label{eq:utildeupd}  
      \enskip . 
\end{eqnarray}
Equation~(\ref{eq:species}) is the species evolution equation, where
$X_k$ is the mass fraction of species $k$, with creation rate
$\omegadot_k$ provided by the nuclear reaction network.  The mass
density, $\rho$, is simply $\rho = \sum_k (\rho X_k)$.  
For the velocity evolution equation, (\ref{eq:utildeupd}), $\pi$ is
the dynamic pressure resulting from the asymptotic expansion of 
the pressure in terms of Mach number.  In paper~III we also evolved the enthalpy,
for the sole purpose of getting the temperature to feed into the reaction network.  
For this paper, we instead define the temperature from $\rho$, $p_0$, and $X_k$.
Our experience has shown that, with the spherical geometry, the discretization
errors are minimized by using the hydrostatic, radial base state pressure to 
define temperature.  We will revisit this in a future paper.
This system of equations
is identical to that presented in paper~III, with $\ubold = \ut$,
$w_0 = 0$, and $\partial p_0/\partial t = 0$.

The velocity field is subject to a constraint equation,
\begin{equation}
\nablab \cdotb (\beta_0 \ubold )  = \beta_0 S
 \enskip ,  \label{eq:tildeconstraint}
\end{equation}
with
\begin{equation}
\label{eq:beta_0}
\beta_0(r) = \rho_0(0) \exp \left ({\int_0^r \frac{1}{\gammabar p_0}
                          \frac{\partial p_0}{\partial r^\prime} \, dr^\prime}
\right )
\enskip ,
\end{equation}
where $\gammabar$ is the average over a layer of $d (\log p)/d (\log
\rho)$ at constant entropy, and
\begin{equation}
  S =  -\sigma  \sum_k  \xi_k \omegadot_k  +
  \frac{1}{\rho p_\rho} \sum_k p_{X_k}  {\omegadot}_k  + \sigma \Hnuc \enskip .
\label{eq:defineS}
\end{equation}
Here, $p_{X_k} \equiv \left . \partial p / \partial X_k \right
|_{\rho,T,X_{j,j\ne k}}$, $\xi_k \equiv \left . \partial h / \partial
X_k \right |_{p,T,X_{j,j\ne k}}$, $p_\rho = \left . \partial
p/\partial \rho \right |_{T, X_k}$, and $\sigma = p_T/(\rho c_p
p_\rho)$, with $p_T \equiv \left . \partial p / \partial T \right
|_{\rho, X_k}$ and $c_p \equiv \left . \partial h / \partial T \right
|_{p,X_k}$ the specific heat at constant pressure.  In these
derivatives, $h$ is the specific enthalpy, defined in terms of 
the specific internal energy, $e$, pressure, and density as $h = e + p/\rho$.  
Finally, $\Hnuc$ is the 
nuclear energy release (with units of erg g$^{-1}$ s$^{-1}$) as computed from our reaction network. 
Physically, $S$
represents the local compressibility effects due to heat release from
reactions and composition changes.  The presence of the density-like quantity $\beta_0$ inside
the divergence in the constraint captures the expansion of a parcel of
fluid as it rises in the hydrostatically stratified star.

We refer the reader to the extensive
comparisons with compressible algorithms in papers I through III that
demonstrate the validity of the low Mach number approximation.  
For the most part, the algorithm to evolve the star follows closely
that described in paper III.  
For the construction of the advective terms, 
the interface states are again
constructed using a piecewise linear unsplit Godunov scheme based on
that of \citet{colella1990}, but we now use
the full corner-coupling scheme developed by
\citet{saltzman1994}.   In the
subsections below we point out the differences for the present
application.

\subsection{\label{sec:mapping} Mapping}

Since the one-dimensional radial base state is not aligned with any
of the axes in the three-dimensional Cartesian grid,
the discretization of 
quantities that involve both the base state and the full state becomes complicated.
Various parts of the algorithm (such as the averaging operations)
require a mapping between the base state and the full state.
Because the base state is not aligned with the Cartesian coordinate axes,
we are free to choose the base state resolution independent of the 
Cartesian grid spacing.  Numerical experimentation has shown that 
setting the base state resolution, $\Delta r,$ to be finer than
the Cartesian grid resolution, $\Delta x,$ gives the best results.
(Here we assume $\Delta x = \Delta y = \Delta z.$)  
For the present simulations, we use $5 \Delta r = \Delta x$.
We refer to the procedure that maps data from 1-d to 3-d as {\tt fill\_3d}, and the
complementary procedure that maps from 3-d to 1-d as {\tt average}.

\begin{figure*}
\begin{center}
\epsscale{0.7}
\plotone{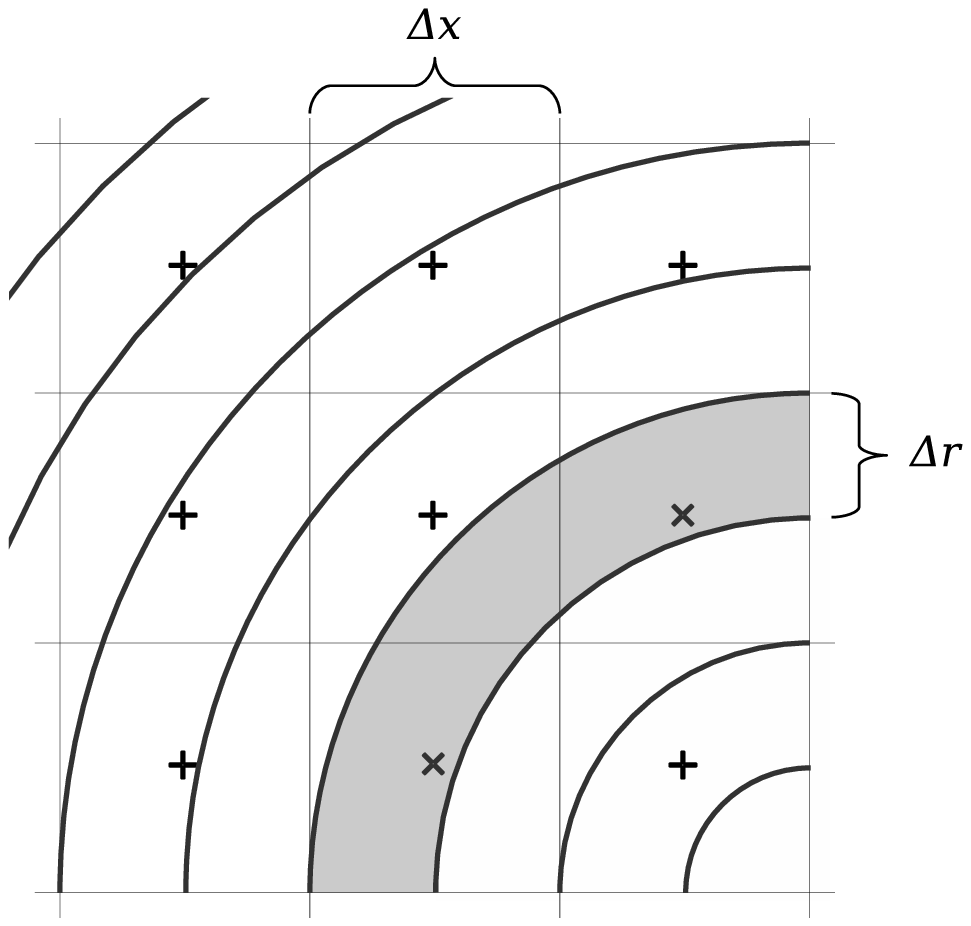}
\epsscale{1.0}
\end{center}
\caption{\label{fig:mapping} The Cartesian grid and spherical base
state (shown here in 2-d for simplicity, using $2 \Delta r = \Delta x$).
Here we represent the
spherical base state as concentric shells (black curved lines).  Since the
base state is not aligned with the Cartesian grid,
we need to map between the two configurations.
The `$+$' symbols represent the Cartesian zone centers.  In our
mapping from the radial profile to the Cartesian grid, 
the zones marked with the `$\times$' symbol are assigned the value 
from the gray-shaded radial bin.}
\end{figure*}

Figure~\ref{fig:mapping} shows the Cartesian grid overlaid by the 
spherical base state (for simplicity, the figure is drawn in 2-d 
using $2 \Delta r = \Delta x$).
The {\tt fill\_3d} procedure computes the distance of the center of
cell indexed by $(i,j,k)$ from the center of the star,
\begin{equation}
r = \sqrt{(x_i - x_c)^2 + (y_j - y_c)^2 + (z_k - z_c)^2} \enskip .
\end{equation}
We use this radius to find the corresponding radial bin as $n = \mathtt{int}(r
/ \Delta r)$ (here, our convention is to use 0-based indexing for the
base state).  We can then initialize a Cartesian cell quantity $q$ from its
corresponding base state quantity, $q_0,$ as $q_{i,j,k} = q_{0,n}$.

For the {\tt average} process, we first define a coarse 1-d radial array 
with $\Delta r_{\rm c} = \Delta x$.  Then, for each cell indexed by
$(i,j,k)$ we again compute the radius, $r,$ as above, and 
define the index of the corresponding coarse radial bin, 
$n_{\rm c} = \mathtt{int}(r / \Delta r_{\rm c})$.  
We define $q_{0,n_{\rm c}}$ as the average of all the $q_{i,j,k}$ whose 
Cartesian cell centers map into coarse radial bin $n_{\rm c}$.
Next, we construct edge-centered states on the coarse radial bin using 
the fourth order approximation,
$q_{0,n_{\rm c}+\myhalf} = (7/12)(q_{0,n_{\rm c}}+q_{0,n_{\rm c}+1}) 
- (1/12)(q_{0,n_{\rm c}-1}+q_{0,n_{\rm c}+2})$.  Finally, for each coarse radial bin
we construct a quadratic profile using $q_{0,n_{\rm c}-\myhalf}, q_{0,n_{\rm c}}$ 
and $q_{0,n_{\rm c}+\myhalf}$.  This is based on the interpolating polynomial used by 
the PPM scheme to find edge states \citep{ppm}.  Specifically, for 
$n_{\rm c}\Delta r_{\rm c} \le r \le (n_{\rm c}+1)\Delta r_{\rm c}$, the 
interpolating polynomial is
\begin{equation}
q_0(r) = q_{0,n_{\rm c}-\myhalf} + \xi(r)\left\{\Delta q_{n_{\rm c}} + q_{6,n_{\rm c}}[1 - \xi(r)]\right\},
\end{equation}
with
\begin{equation}
\xi(r) = \frac{r - n_{\rm c}\Delta r_{\rm c}}{\Delta r_{\rm c}},
\end{equation}
\begin{equation}
\Delta q_{n_{\rm c}} = q_{0,n_{\rm c}+\myhalf} - q_{0,n_{\rm c}-\myhalf},
\end{equation}
and
\begin{equation}
q_{6,n_{\rm c}} = 6\left[q_{0,n_{\rm c}} - 
\frac{1}{2}\left(q_{0,n_{\rm c}+\myhalf} + q_{0,n_{\rm c}-\myhalf}\right)\right].
\end{equation}
We note that since we are not evolving the base state in the simulation
presented here, the feedback from the full state to the base state through
{\tt average} is limited to computing $\gammabar$, as needed for updating
$\beta_0$.

\subsection{Microphysics}

We use the general stellar equation of state described by
\citet{timmes_swesty:2000,flash}, which includes contributions from
electrons, ions, and radiation.  For these calculations we include the
effects of Coulomb corrections included in the publicly available
version of this EOS \citep{timmes_eos}.

Our reaction network is unchanged from paper~III, and is a single-step
$^{12}\mathrm{C} + \, ^{12}\mathrm{C}$ reaction using screening as
described in
\citet{graboske:1973,weaver:1978,alastuey:1978,itoh:1979}, resulting
in $^{24}\mathrm{Mg}$ ash.  We release the energy corresponding to the
binding energy difference between the magnesium ash and carbon fuel.
Paper~III provides full details on how the reaction network is
solved.  Our only change from the implementation there is that we
now update the temperature at the end of the reaction step. 
Finally we note that we
do not call the reaction network for densities below $\rhocutoff$.

We note that by integrating the reaction rate equation, we are dealing
with reactions differently than \citet{kuhlen-ignition:2005}.  There,
an analytic approximation to the reaction rate was used and evaluated
given a temperature and density.  Our method extends more easily to a
full reaction network.  A second difference is that
\citet{kuhlen-ignition:2005} burned to a mix of neon and magnesium,
leading to a slightly lower energy release.  This difference may
affect the timescales we see in the calculation, but we don't expect
it to introduce qualitative differences.

\subsection{Initial Model}
\label{sec:initial_model}

We begin with an initial 1-d white dwarf model produced with the
stellar evolution code, Kepler \citep{weaver:1978}.  This model was
evolved to the point where the central temperature is $6\times 10^9$~K
and the central density is $2.6\times 10^9~\gcc$.  The composition is
about half $^{12}\mathrm{C}$ and half $^{16}\mathrm{O}$, with a small
amount ($<$ 0.5\%) of ash in the center of the star.  The total mass
of the star is $1.382~M_\odot$.

We follow the procedure outlined in \cite{ppm-hse} to convert the
initial model from the one-dimensional Lagrangian mesh used by Kepler
to the uniformly-zoned Eulerian grid used in our calculation.  It is
important that the initial model satisfy hydrostatic equilibrium
discretely with our equation of state on the base state grid we use
for our simulation.  In particular, we want to enforce the following
discretization of hydrostatic equilibrium:
\begin{equation}
p_{0,i+1} - p_{0,i} = \frac{1}{2} \Delta r ( \rho_{0,i} + \rho_{0,i+1} ) g_{i+\myhalf} 
\label{eq:hse}
\enskip .
\end{equation}
Hydrostatic equilibrium alone does not specify our initial model, since we must also specify
the initial temperature.  In the interior of the star, where convection
dominates, constant entropy is a good approximation.
We use this constraint together with equation (\ref{eq:hse}) and the equation
of state to find the temperature, density, and pressure throughout the
inner region of the star.  For the composition, we use the profile
provided by the Kepler model, but, since we are using a reduced
network, we group together the $^{20}$Ne and $^{24}$Mg ash into a
single composition variable.

\begin{figure*}
\begin{center}
\epsscale{0.6}
\plotone{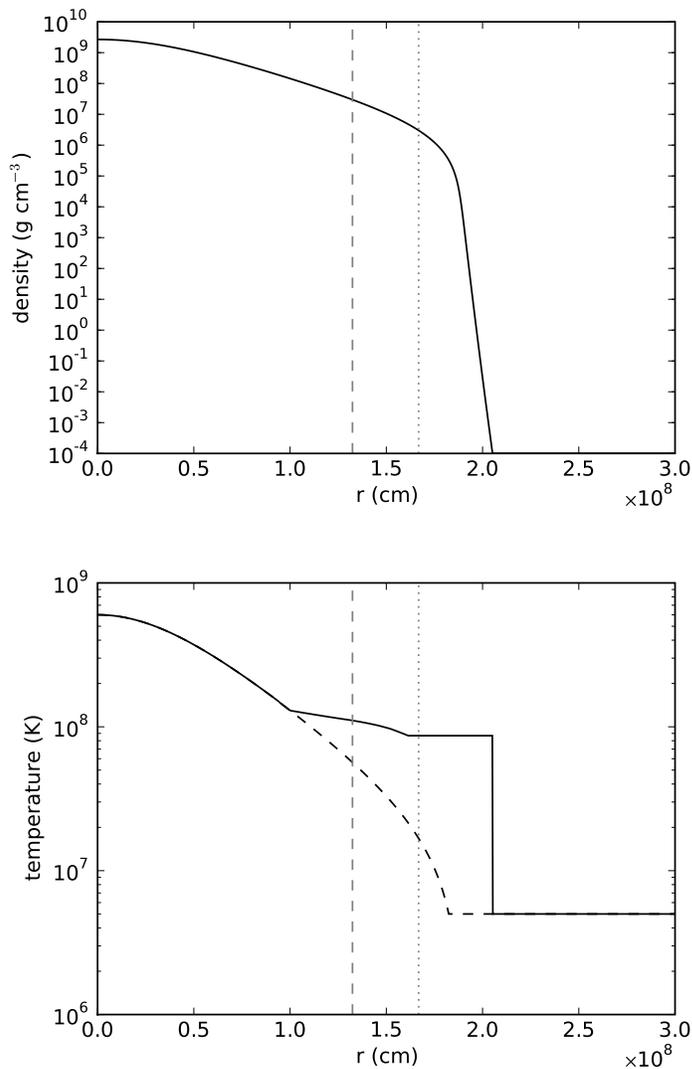}
\epsscale{1.0}
\end{center}
\caption{\label{fig:initial_model} The initial model used for the full
  star convection calculation.  The top panel shows the density and the
  bottom panel shows the temperature.  In both panels, the vertical
  dotted gray line represents the location of the low density
  cutoff---data outside of this cutoff are not used by our
  calculations.  The vertical dashed gray line indicates where our
  sponge forcing term begins.  For the temperature plot, the solid
  line represents the initial model used in our calculation and the
  dashed line represents the temperature structure for a completely
  isentropic model.}
\end{figure*}

The convective region is surrounded by an outer, convectively stable
region.  When the isentropic temperature profile drops below the
temperature provided by the Kepler model, we switch to using the
Kepler temperature.  Figure~\ref{fig:initial_model} shows our final
temperature profile, along with a completely isentropic model for
reference.  The departure of the two temperature curves marks the
boundary of the convective region.  The mass of the inner isentropic
region of the star is $1.131~M_\odot$.  We note that the spatial extent 
of the convective zone in the white dwarf is somewhat
uncertain.  Different assumptions about the accretion history of the white
dwarf would lead to different mass convection zones.

Overall, this procedure results in a slight adjustment of the
structure of the star compared to the initial Kepler model.  The
resulting model serves as the initial base state for our calculation.
As discussed in papers II and III, outside of the star we cannot
bring the density down to arbitrarily small values, as that would 
result in too high a velocity (a consequence of our constraint equation).  
In practice, we impose a cutoff at a moderately small density,
$\rhocutoff$, and set the density to this constant value outside of
the star.  For the main calculation presented here, we choose
$\rhocutoff = 3\times 10^6~\gcc$.  While this may sound high, we note
that the mass of the star enclosed by $\rhocutoff$ is
$1.378~M_\odot$---a 0.2\% difference from the total mass of the star.
We note also that as in paper~III, we use an anelastic cutoff, the
density below which the coefficient, $\beta_0,$ of our velocity constraint
is defined by keeping $\beta_0/ \rho_0$ constant.  In
this paper, we always set the anelastic cutoff to be $\rhocutoff$.

The initial three-dimensional state is set by using the {\tt fill\_3d} 
routine in \S~\ref{sec:mapping} to interpolate $\rho_0$, $p_0$, 
$X_{k,0}$, and $T_0$ to each cell center.   The initial velocity field
is not as well defined.
The one-dimensional stellar evolution code used mixing length theory
to describe convective mixing in the interior of the star.  When we
map the model onto our three-dimensional grid, there is a region that
is convectively unstable (corresponding to the region in
Figure~\ref{fig:initial_model} 
where $r < 1.0\times 10^8~\mathrm{cm}$).  However, there
is not enough information in the one-dimensional model to initialize a
three-dimensional velocity field that correctly represents the
convective field.  

If we start with zero initial velocity, then at $t=0$ the
reactions near the core generate a large amount of energy and the highly
nonlinear form of the reaction rate means that the energy release
quickly grows.  Without an initial velocity field to advect some 
of this energy away from the core, the energy generation grows too
quickly, and an unphysical runaway occurs. However, by starting
with an initial non-zero velocity field, our simulation very quickly finds 
a convective velocity field that balances the energy generation at the core.  
Thus we define a set of Fourier modes,
\begin{mathletters}
\begin{eqnarray}
C^{(x)}_{l,m,n} &=& \cos \left (\frac{2 \pi l x}{\sigma} + \phi^{(x)}_{l,m,n} \right ) \\
C^{(y)}_{l,m,n} &=& \cos \left (\frac{2 \pi m y}{\sigma} + \phi^{(y)}_{l,m,n} \right ) \\
C^{(z)}_{l,m,n} &=& \cos \left (\frac{2 \pi n z}{\sigma} + \phi^{(z)}_{l,m,n} \right )
\end{eqnarray}
\end{mathletters}
and
\begin{mathletters}
\begin{eqnarray}
S^{(x)}_{l,m,n} &=& \sin \left (\frac{2 \pi l x}{\sigma} + \phi^{(x)}_{l,m,n} \right ) \\
S^{(y)}_{l,m,n} &=& \sin \left (\frac{2 \pi m y}{\sigma} + \phi^{(y)}_{l,m,n} \right ) \\
S^{(z)}_{l,m,n} &=& \sin \left (\frac{2 \pi n z}{\sigma} + \phi^{(z)}_{l,m,n} \right )
\end{eqnarray}
\end{mathletters}
where $\sigma$ is the characteristic scale of the perturbation and the
$\phi^{\{x,y,z\}}_{l,m,n}$ are randomly generated phases between [0,
  $2\pi$].  We then compute the total contribution to the velocity
perturbation from the modes as
\begin{mathletters}
\begin{eqnarray}
u' &=& \sum_{l=1}^3 \sum_{m=1}^3 \sum_{n=1}^3 \frac{1}{N_{l,m,n}} 
   \left [ - \gamma_{l,m,n} m C^{(x)}_{l,m,n} C^{(z)}_{l,m,n} S^{(y)}_{l,m,n}
           + \beta_{l,m,n}  n C^{(x)}_{l,m,n} C^{(y)}_{l,m,n} S^{(z)}_{l,m,n} \right ] \\
v' &=& \sum_{l=1}^3 \sum_{m=1}^3 \sum_{n=1}^3 \frac{1}{N_{l,m,n}} 
   \left [\phantom{+}  \gamma_{l,m,n} l C^{(y)}_{l,m,n} C^{(z)}_{l,m,n} S^{(x)}_{l,m,n}
           - \alpha_{l,m,n} n C^{(x)}_{l,m,n} C^{(y)}_{l,m,n} S^{(z)}_{l,m,n} \right ] \\
w' &=& \sum_{l=1}^3 \sum_{m=1}^3 \sum_{n=1}^3 \frac{1}{N_{l,m,n}} 
   \left [ - \beta_{l,m,n}  l C^{(y)}_{l,m,n} C^{(z)}_{l,m,n} S^{(x)}_{l,m,n}
           + \alpha_{l,m,n} m C^{(x)}_{l,m,n} C^{(z)}_{l,m,n} S^{(y)}_{l,m,n} \right ]
\end{eqnarray}
\end{mathletters}
where $\alpha_{l,m,n}$, $\beta_{l,m,n}$, and $\gamma_{l,m,n}$ are
randomly generated amplitudes between [-1, 1], and $N_{l,m,n} =
\sqrt{l^2 + m^2 + n^2}$ is the normalization,

A perturbational velocity field is then computed as
\begin{mathletters}
\label{eq:velpert}
\begin{eqnarray}
u'' &=& \frac{A u' }{2} \left [ 1 + \tanh \left ( \frac{r_\mathrm{pert} - r}{d} \right ) \right ]\\
v'' &=& \frac{A v' }{2} \left [ 1 + \tanh \left ( \frac{r_\mathrm{pert} - r}{d} \right ) \right ]\\
w'' &=& \frac{A w' }{2} \left [ 1 + \tanh \left ( \frac{r_\mathrm{pert} - r}{d} \right ) \right ]
\end{eqnarray}
\end{mathletters}
where the $\tanh$ profile gradually cuts off the perturbation at a
radius $r_\mathrm{pert}$ with a transition thickness $d$.  Finally,
the initial velocity field is computed by applying the projection to
$(u'',v'',w'')$ to ensure that it satisfies the divergence constraint.
We pick the amplitude, $A$, to be small, and independent of the
velocities used in the 1-d stellar evolution model.  Once the flow
field is established, we expect the details of the initial velocity
field to be forgotten.  This is an area we will explore in a
subsequent paper. 

Throughout the calculation we solve the reaction network to compute the 
energy release that drives the convection.  By starting at
a low initial central temperature, we thus expect a realistic flow
field to build up over time as the central temperature increases
from the reactions.
In this respect we differ from the initialization procedure used in
\citet{kuhlen-ignition:2005}.  In their anelastic approximation, they
carried the perturbational temperature separately from the base state
temperature, and to initialize the flow field they evaluated the
carbon burning heating term using only the base state temperature.  By
excluding the perturbational temperature, they left out the nonlinear
feedback in the extremely temperature-sensitive carbon reaction
rate, and therefore built a flow field without the chance of runaway.  
Once the flow field was established, they fed the temperature
perturbations back into the reaction rate to watch the runaway.

\subsection{Sponging}

\begin{figure*}
\begin{center}
\epsscale{0.7}
\plotone{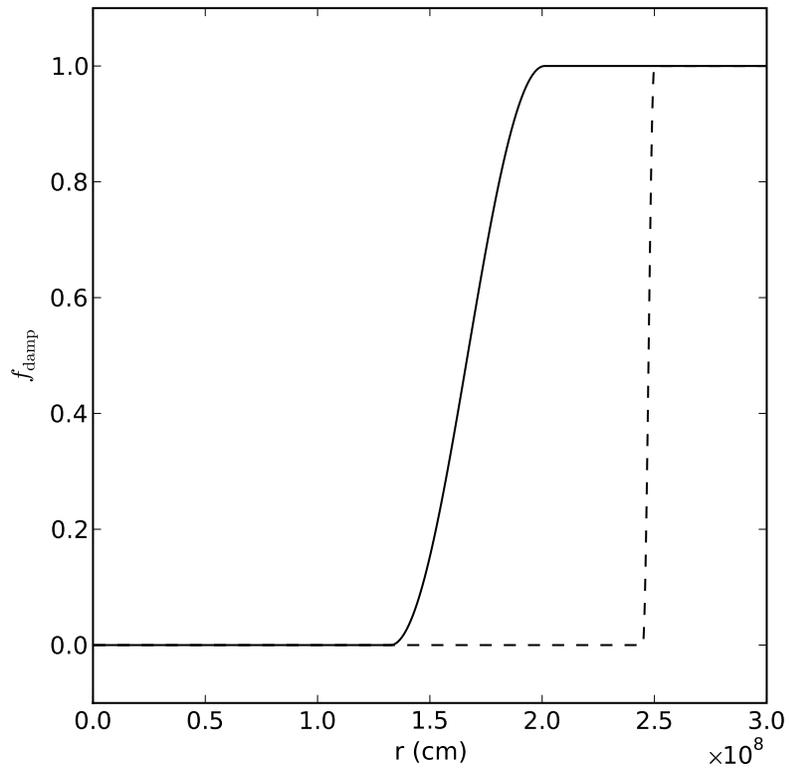}
\epsscale{1.0}
\end{center}
\caption{\label{fig:sponge} The inner sponge function, $f_\mathrm{damp},$
as a function of radius for $\rhocutoff = 3\times 10^6~\gcc$ (solid line),
and the outer sponge function for $D = 5\times 10^8~\mathrm{cm}$ and a $384^3$
grid (dashed).}
\end{figure*}

As described in paper~III, we use a sponge to damp the velocities outside
of our region of interest.  We use the same functional form here, with 
the velocity forcing given by
\begin{equation}
\ubold^\mathrm{new} = \ubold^\mathrm{old} - \dt\, \kappa f_\mathrm{damp} \ubold^\mathrm{new} \enskip ,
\end{equation}
where $\kappa$ is a frequency.  For all results presented here, we use
$\kappa = 10~\mathrm{s}^{-1}$.
The sponge factor has the form:
\begin{equation}
f_\mathrm{damp} = \left \{ \begin{array}{ll}
   0                                & \mbox{if $r < r_\mathrm{sp}$} \\  \noalign{\medskip}
  \displaystyle \frac{1}{2} \left \{ 1 -
           \cos \left [ \pi \left ( \frac{r - r_\mathrm{sp}}
                                         {r_\mathrm{tp} - r_\mathrm{sp}}
                             \right )\right ] \right \}
                                    & \mbox{if $r_\mathrm{sp} \le r < r_\mathrm{tp}$} \\  \noalign{\medskip}
   1                                & \mbox{if $r \ge r_\mathrm{tp}$} \end{array} \right .\enskip.
\end{equation}
The quantity $r_\mathrm{sp}$ represents the radius where the sponging
term gradually begins to turn on, and is set to the radius
corresponding to 10$\cdot \rhocutoff$.  The top of the sponge,
$r_\mathrm{tp}$, where the sponge is in full effect is set as
$r_\mathrm{tp} = 2 r_\mathrm{md} - r_\mathrm{sp}$, with
$r_\mathrm{md}$ set to the radius corresponding to the $\rhocutoff$.
As noted above, we use a $\rhocutoff = 3\times
10^6~\gcc$ for these calculations, so the corresponding density where
our sponging begins is $3\times 10^7~\gcc$.  Based on our initial
model, $1.320~M_\odot$ of the star is contained within
$r_\mathrm{sp}$---the sponge only affects the very outer portion of
the star.  Figure~\ref{fig:initial_model} shows the location of
$r_\mathrm{sp}$ for our initial model---we see that it is well outside
the convectively unstable region.  Figure~\ref{fig:sponge} shows
$f_\mathrm{damp}$ vs.\ $r$ for $\rhocutoff = 3\times 10^6~\gcc$.

This sponge is effective in damping the velocities at the edge of the
star.  Our domain is $D = 5\times 10^8$~cm on a side, so the distance
from the center of the star along one of the coordinate axes to the
edge of the domain is $2.5\times 10^8$~cm.  The distance from the
center to a corner of the domain is $\sqrt{3}$ larger.  Because we are
placing a spherical star in a cubic domain, we found that we need an
additional sponge to damp the velocities in the outer corners of the
domain---well outside of the star.  We define an outer sponge of the
same form as above, but with $r_\mathrm{tp} = D/2$ and $r_\mathrm{sp}
= r_\mathrm{tp} - 4 \Delta x$, where $\Delta x$ is the grid spacing,
and $\kappa$ set to $10$ times the value of the inner sponge.
This additional sponge is included in the momentum equation in the same
fashion as the inner sponge.  Figure~\ref{fig:sponge} shows the profile
of this additional sponge as well.


\section{Results}

Our main goal in these simulations is to study the convection
in the white dwarf up to the point of ignition.  In this section we
present results for our main $384^3$ convection calculation, 
supporting calculations with lower resolution, as well as a test problem.
In each case, the code was run with an advective CFL number of 0.5 with 
the star centered in a domain $5\times 10^8~\mathrm{cm}$ on a side.

\subsection{Test Problem: Isentropically Stratified Star}

To test the interaction between the spherical base state and the 3-d
Cartesian representation of the star, we perform a simple advection test with 
an analytic solution.  First,  we construct a completely isentropic
initial model.  This is achieved by picking a central density of $2.6\times 10^9~\gcc$ and a
central temperature of $6\times 10^8~\mathrm{K}$, and a uniform
composition of 0.3 $^{12}\mathrm{C}$ and 0.7 $^{16}\mathrm{O}$, and
integrating outward using our hydrostatic equilibrium constraint,
equation~(\ref{eq:hse}), and forcing the entropy to be constant through
the equation of state.  We initialize the full state using the isentropic
base state with no perturbations.  We also set $\beta_0 = \rho_0$ discretely
(which is true analytically for an isentropic base state and constant $\gammabar$), and disable all reactions 
and heating.  The constraint is now identical to the anelastic constraint, 
$\nablab \cdot (\rho_0 U) = 0$.  

Under these conditions, the continuity equation becomes:
\begin{equation}
\frac{\partial \rho}{\partial t} = -\nablab \cdot (\rho U) = -\nablab \cdot (\rho_0 U) = 0 \enskip ,
\end{equation}
using the fact that $\rho = \rho_0$ initially, and the 
anelastic constraint.  As a result, we see that the density should
remain constant in the star regardless of the velocity field.

This provides a means to test our mapping procedure.  If we start with
an isentropically stratified star and seed a random velocity field, the
density should not change with time.  For our test, we
start with a random velocity field described by equation (\ref{eq:velpert}).
For the amplitude of the perturbation, we set $A = 10^7~\cms$---this
is typical of the highest velocities we expect to see in our
convection calculations.  For the size of the perturbation, we set
$r_\mathrm{pert} = 5\times 10^7~\mathrm{cm}$---this value represents
about half the size of the expected convective region in the white
dwarf.  Finally, we set the characteristic wavelength of the
perturbation, $\sigma = 10^7~\mathrm{cm}$.  We make the transition
between the perturbation and the ambient star sharp, effectively
smaller than our grid resolution, setting $d = 10^5~\mathrm{cm}$.
The resolution is $384^3$, the same as that used in the 
main calculation in the next section.

To assess the change in density with time, we will look at the average
density, $\langle \rho \rangle,$ as a function of radius, and the
deviation of the density as a function of radius, $\delta \rho$.  We
define these as
\begin{equation}
\langle \rho \rangle_r = \frac{1}{{N_\Omega}_r} \sum_{\Omega_r} \rho \enskip ,
\end{equation}
where $\Omega_r$ is the set of cells in the computational domain whose center
falls within the radial bin at radius $r$, and ${N_\Omega}_r$ is the number of cells
in $\Omega_r$.  The RMS fluctuations are
\begin{equation}
\label{eq:delta_rho}
(\delta \rho)_r = \left [ \frac{1}{{N_\Omega}_r} \sum_{\Omega_r} (\rho - \rho_0)^2  \right ]^{\half} \enskip .
\end{equation}
Here we recognize that the base state density, $\rho_0$ represents
the average density at a given radius.  We compute and store $(\rho -
\rho_0)$ for every zone in our computational domain directly in the
code as the simulation runs, and then compute $(\delta \rho)_r$ using
equation (\ref{eq:delta_rho}) with a radial bin spacing $\Delta r =
\Delta x$---this ensures that no interpolation is needed to fill
radial cells.

\begin{figure*}
\begin{center}
\epsscale{1.0}
\plotone{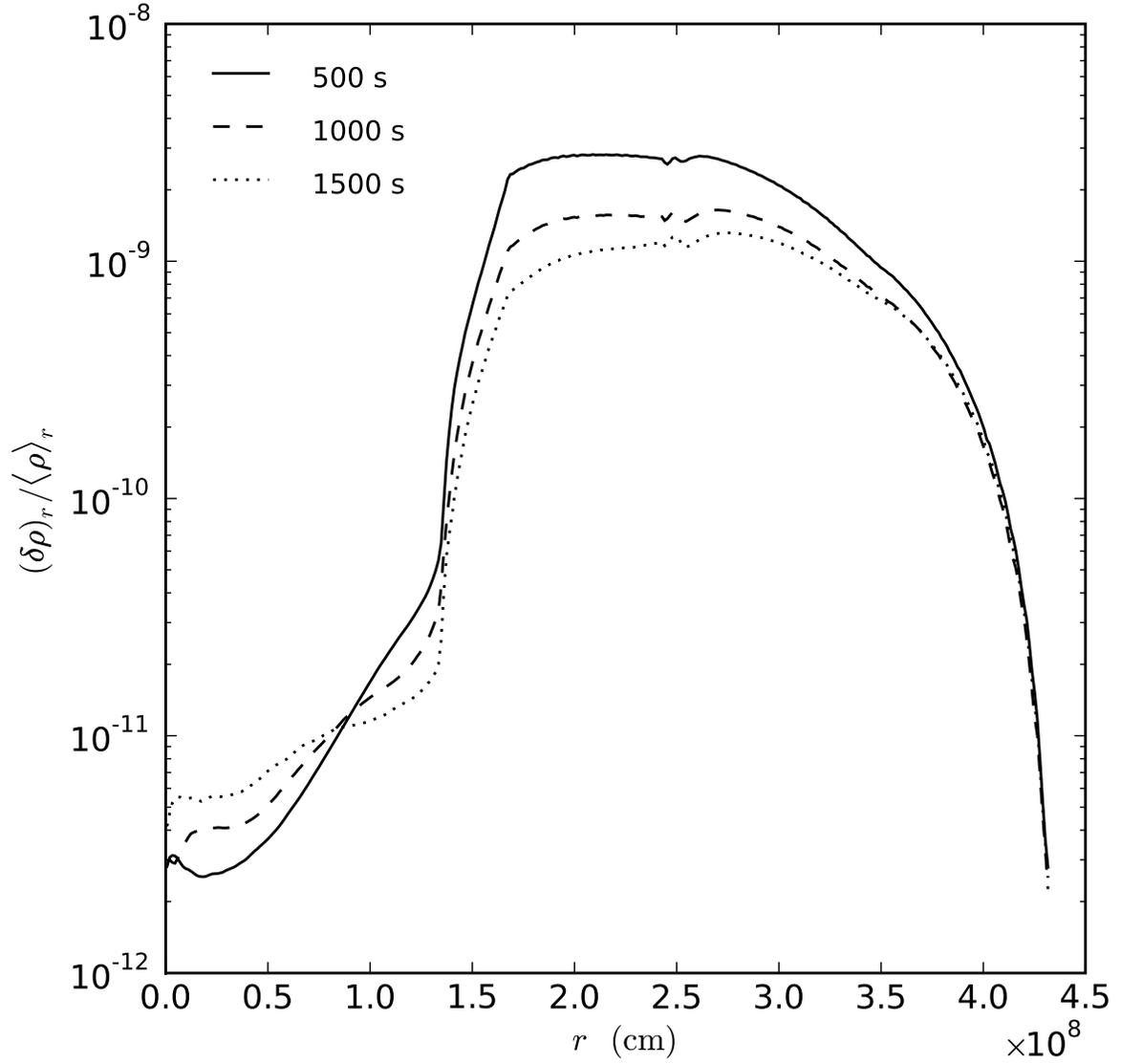}
\epsscale{1.0}
\end{center}
\caption{\label{fig:test_spherical} $(\delta \rho)_r / \langle \rho
  \rangle_r$ vs.\ $r$ for the test problem at 3 times.  We see that the relative
  change in density resulting from a large amplitude velocity perturbation
  is small.}
\end{figure*}

Figure~\ref{fig:test_spherical} shows a plot of $(\delta \rho)_r /
\langle \rho \rangle_r$ vs.\ $r$ at several times.  By normalizing to
the average density, $\langle \rho \rangle_r$, we are seeing a measure
of the relative error in the density from our advection scheme.  As
the plot shows, even after 1500~s of evolution, the error at the
center of the star is $< 10^{-11}$.  Once we are outside of the star
$r > 2\times 10^{8}~\mathrm{cm}$, the error rises, but still stays
below $5\times 10^{-9}$ everywhere.  This demonstrates that our
algorithm accurately preserves $\partial \rho / \partial t = 0$ in the
limiting case of an isentropic model with no heating and $\beta_0 =
\rho_0$.

\subsection{Convection in a White Dwarf}

We model convection in the white dwarf by mapping the initial model
described in \S~\ref{sec:initial_model} onto our Cartesian grid.  
For the initial velocity field, we use $A = 10^5~\cms$, $d = 10^5~\mathrm{cm}$, 
$r_\mathrm{pert} = 2\times 10^7~\mathrm{cm}$, and $\sigma = 10^7~\mathrm{cm}$.

\subsubsection{Diagnostics}

To help us understand the character of the flow in our calculations we
make use of several diagnostic quantities.  We define the region
of interest of the domain, $\Omega_\mathrm{star}$, to be those
computational cells with $\rho > \rhocutoff$, where we have used
$\rhocutoff = 3 \times 10^6~\gcc$ unless otherwise specified.  The
diagnostics defined below are computed every time step, as the code is
running.

The peak temperature in the domain is simply
\begin{equation}
T_\mathrm{peak} = \max_{\Omega_\mathrm{star}} \{ T \}\enskip .
\end{equation}
As the temperature in the star increases considerably toward the
center of the star, we expect the peak temperature to be close to (but
not exactly equal to) the central temperature.

Motivated by previous results that suggest a dipole nature to the flow
\citep{kuhlen-ignition:2005}, we look at several diagnostics based on
the radial fluid velocity.  First we define the radial velocity to be
$v_r = \ubold \cdotb \er$.  Then we compute components of the density-weighted
average radial
velocity in each coordinate direction,
\begin{equation}
\favgradvel_x = 
\sum_{N_{\Omega_\mathrm{star}}} \rho v_r \left ( \frac{x - x_c}{r}\right ) / 
\sum_{N_{\Omega_\mathrm{star}}} \rho \enskip ,
\end{equation}
where $r$ is the distance of a given zone from the center of the star,
and $N_{\Omega_\mathrm{star}}$ is the number of computational zones
contained in the domain $\Omega_\mathrm{star}$.  We similarly compute
$\favgradvel_y$ and $\favgradvel_z$ using the $y$ and $z$ coordinates
and centers.  The relative magnitudes of the components of
$\favgradvel_i$ tell us about the direction of any dipole-nature to
the flow.  In particular, we can derive the directional angles $\phi$
in the $x$-$y$ plane, and $\theta$ as measured from the $z$-axis as
\begin{equation}
\phi = \tan^{-1} \left ( \frac {\favgradvel_y}{\favgradvel_x} \right )
\end{equation}
and
\begin{equation}
\theta = \tan^{-1} \left ( \frac{\sqrt{\favgradvel_x^2 + \favgradvel_y^2}}{\favgradvel_z} \right ) \enskip .
\end{equation}
We could have instead computed the average radial velocity without a
density weighting, but because we are summing over the entire star
(where $\rho > \rhocutoff$), we are including the outer convectively
stable region in the average, where we do not expect to see much
influence from the dipole.  By density-weighting, we are giving
more weight to the center of the star, where the convective
pattern dominates.

Finally, to get a sense of scale for the radial velocity in the 
star, we compute
\begin{equation}
(v_r)_\mathrm{peak} = \max_{\Omega_\mathrm{star}} \{ | v_r | \} \enskip .
\end{equation}

\subsubsection{Long Term Convective Behavior}

Our main result is a $384^3$ calculation of convection in a white
dwarf, starting from an initial model with a central temperature of
$6\times 10^8$~K.  Our computational domain is $5\times 10^8$~cm on a
side, giving us 13~km zones.  Our goal is to follow the convection as
reactions bring the central temperature up over $7\times 10^8$~K, and
into the regime of ignition.

\begin{figure*}
\begin{center}
\mbox{\includegraphics[width=2.1in]{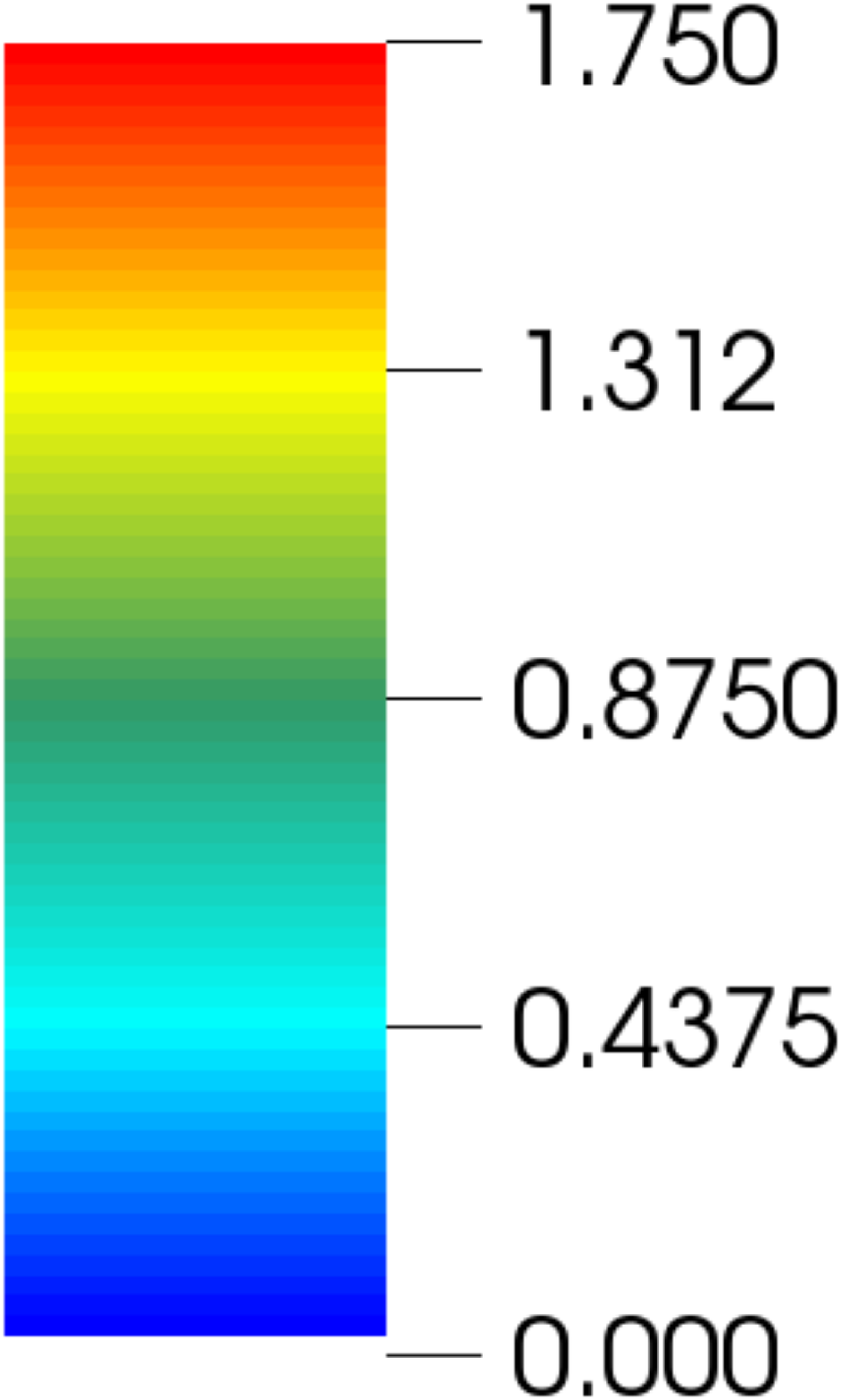}
      \includegraphics[width=2.1in]{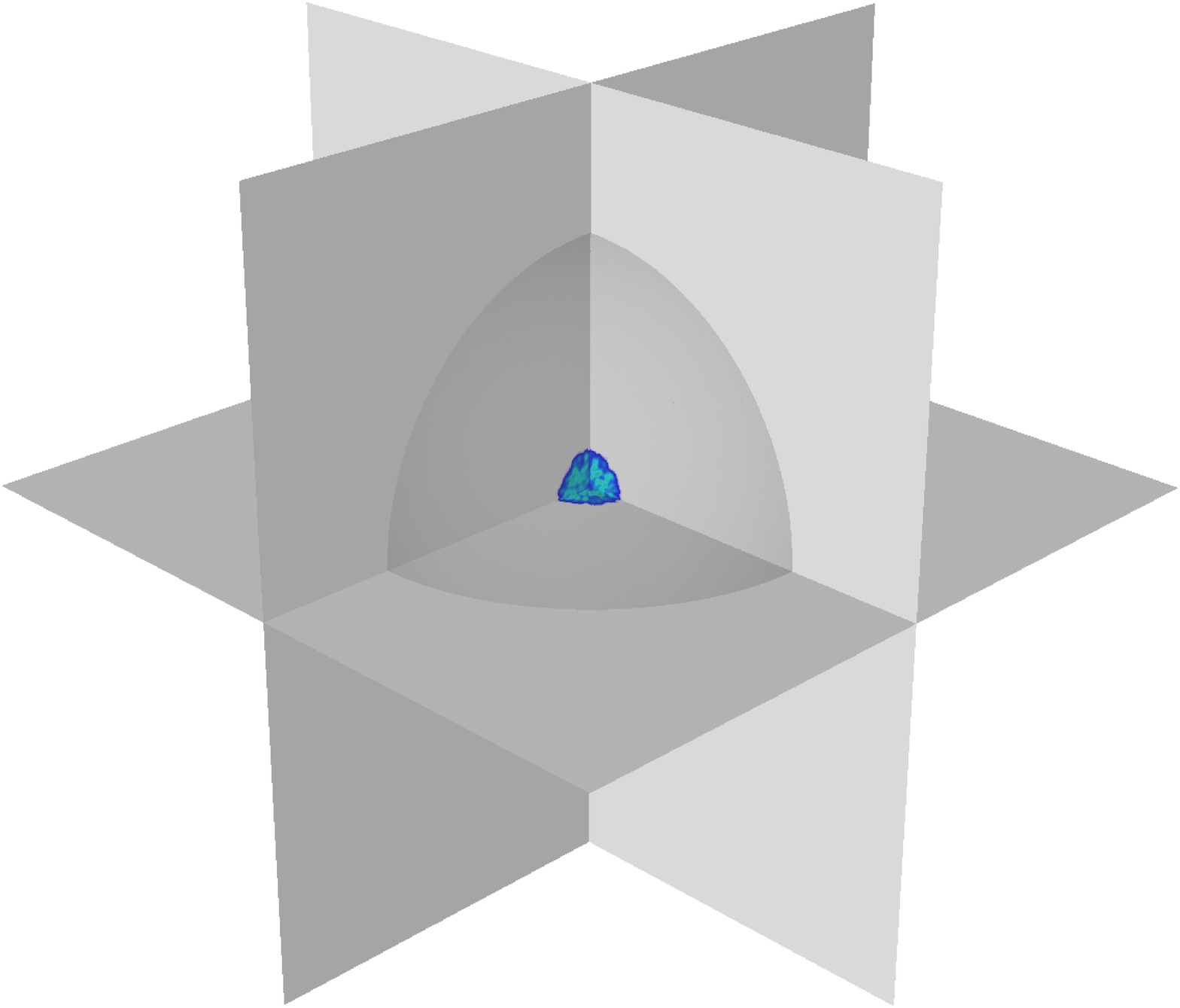}
      \includegraphics[width=2.1in]{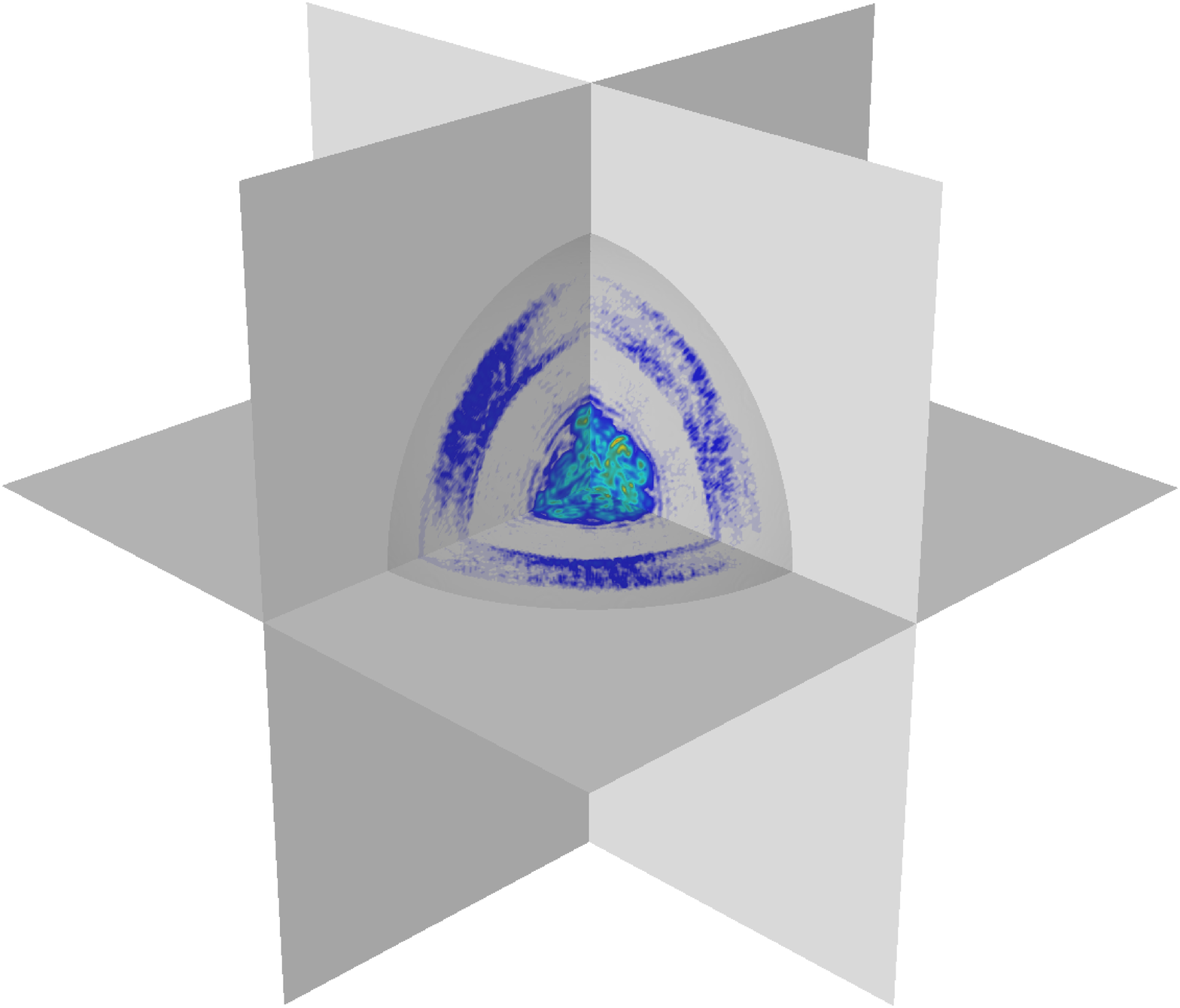}}
\mbox{\includegraphics[width=2.1in]{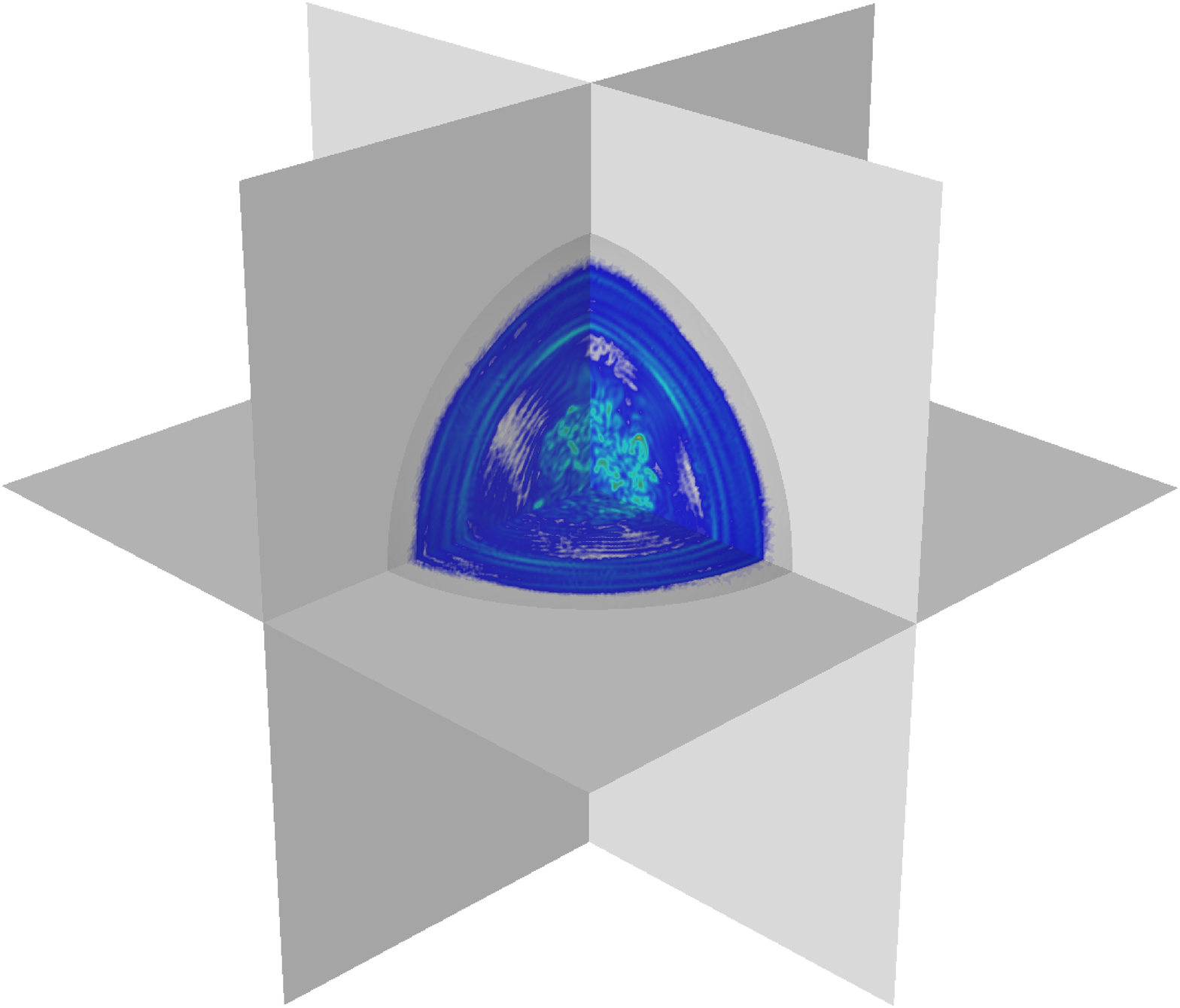}
      \includegraphics[width=2.1in]{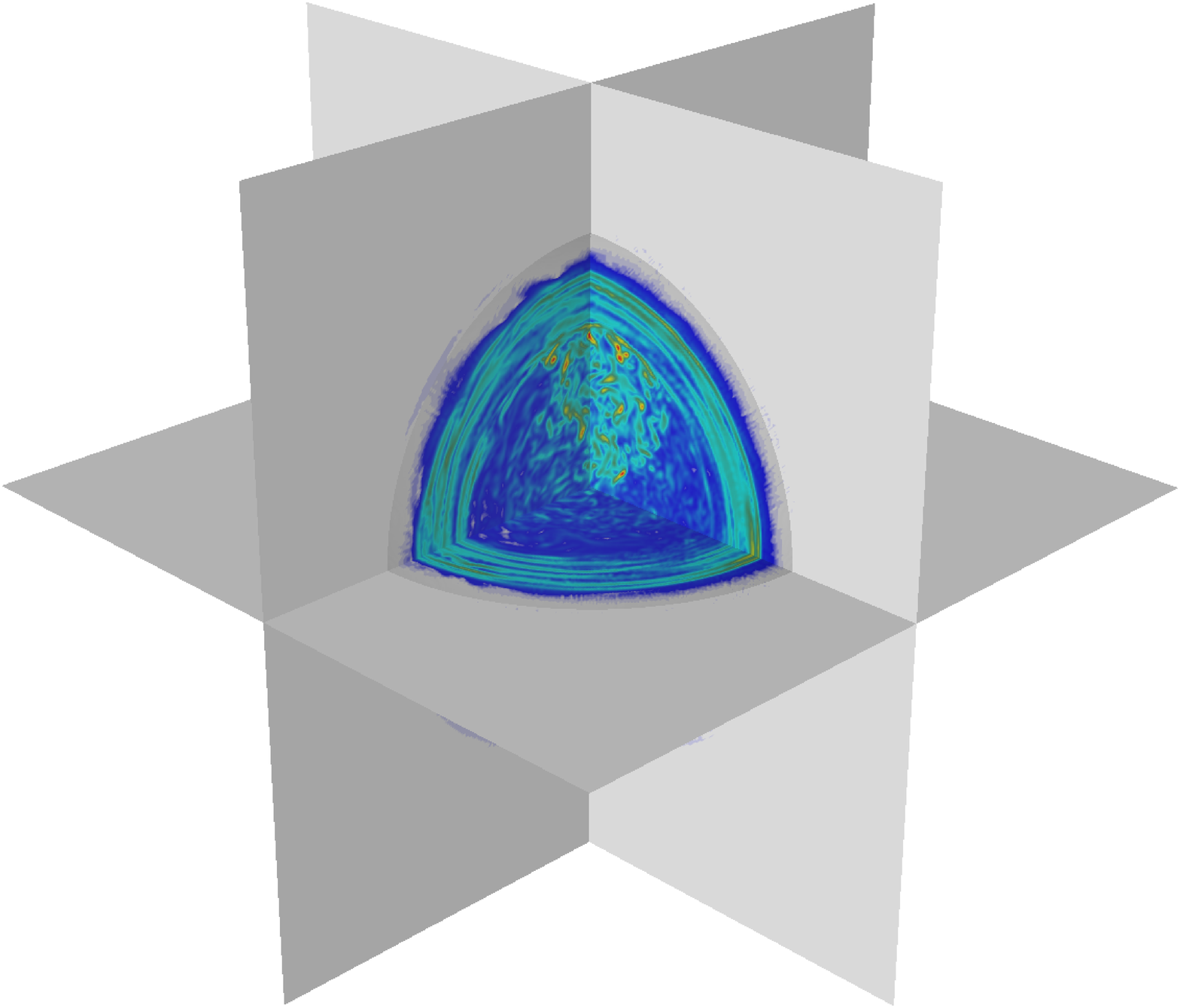}
      \includegraphics[width=2.1in]{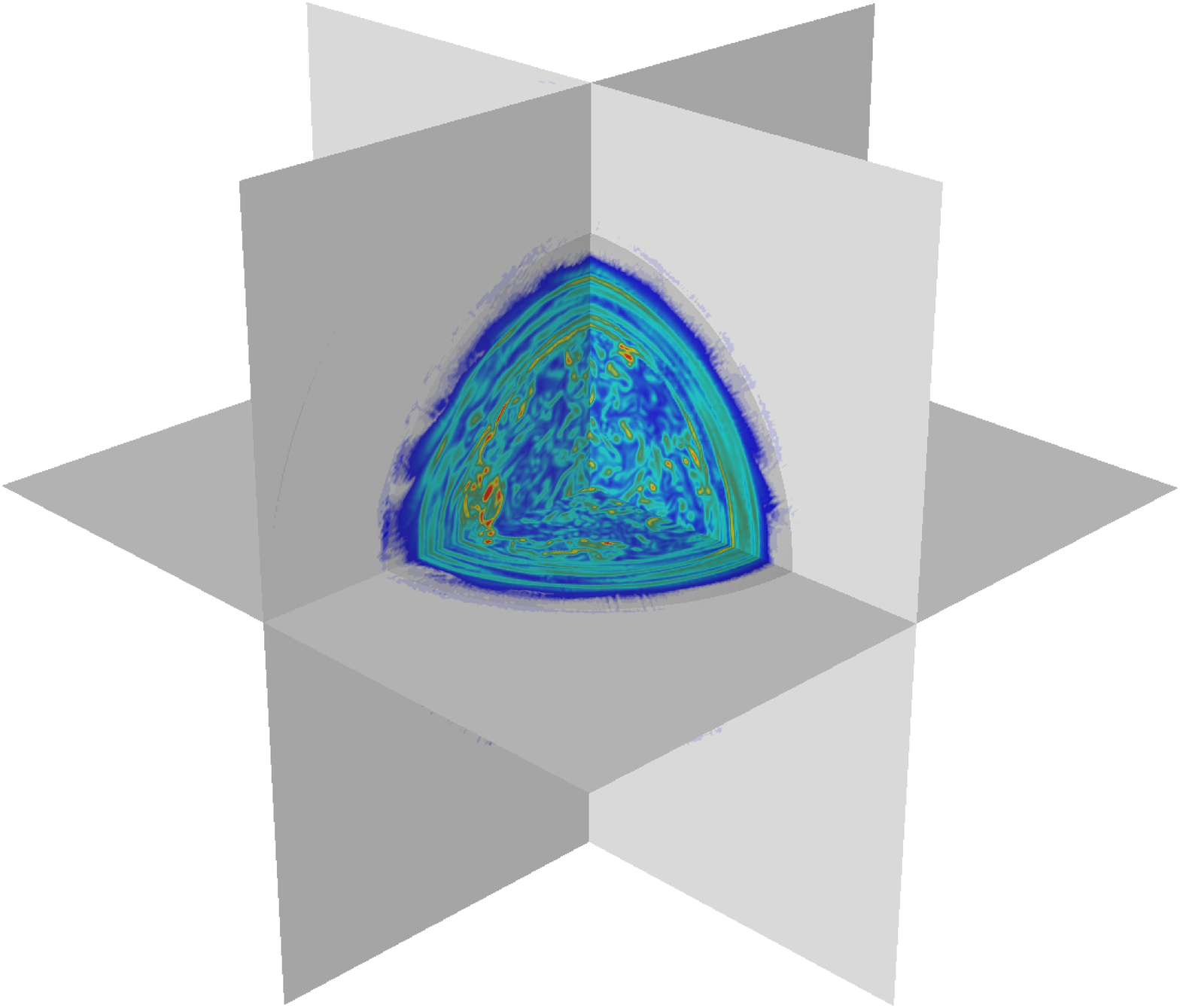}}
\mbox{\includegraphics[width=2.1in]{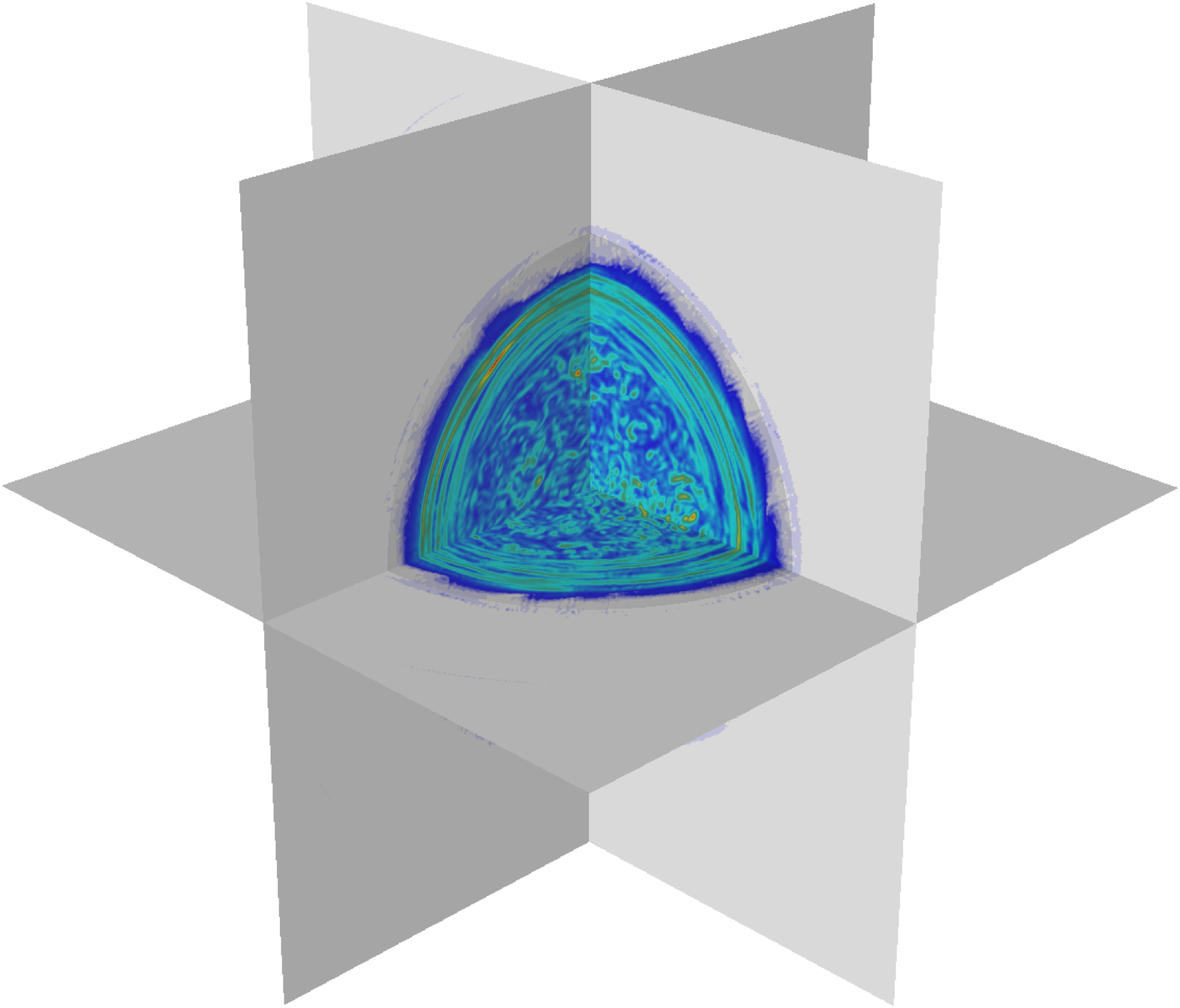}
      \includegraphics[width=2.1in]{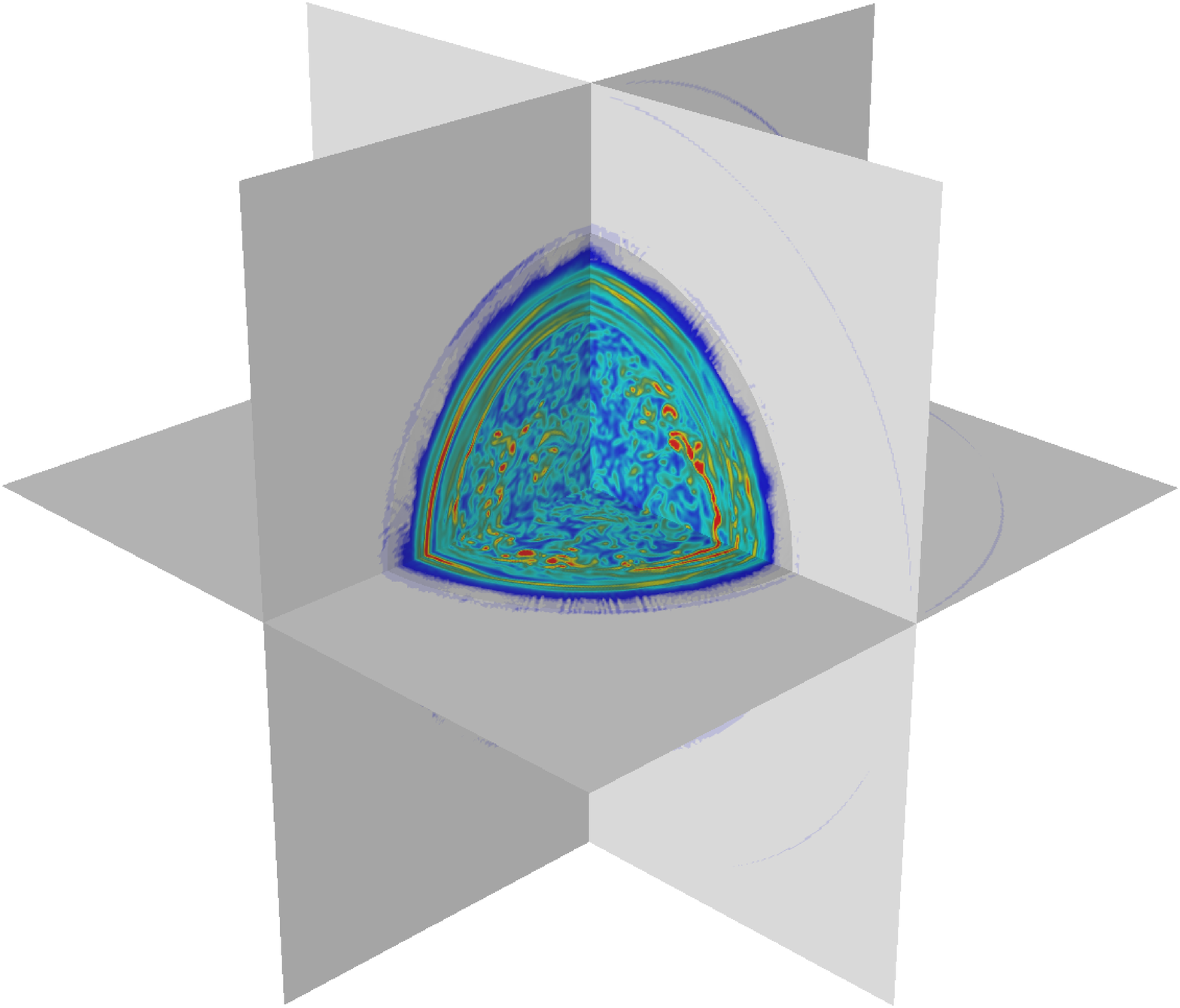}
      \includegraphics[width=2.1in]{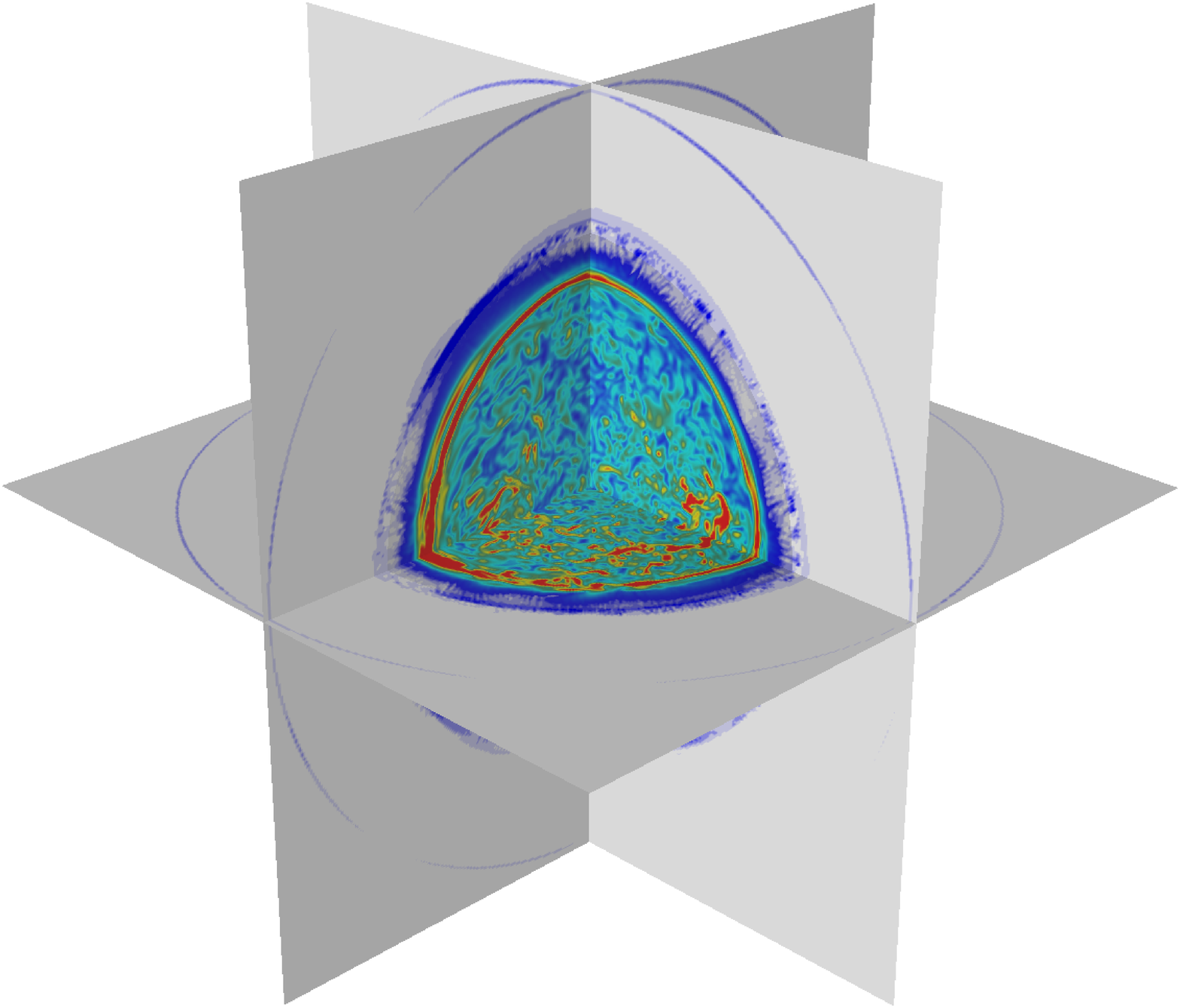}}
\end{center}
\caption{\label{fig:vort} Development of the convective flow in the 384$^3$ calculation.
Here we plot vorticity.  The data scale is capped at 1.75$~s^{-1}$, even though the maximum
vorticity steadily climbs as the simulation progresses, reaching over 13$~s^{-1}$ by the last
panel.  From left to right, top to bottom, the panels show the vorticity 
at 50, 100, 200, 400, 800, 1600, 3200, and 6400~s.}
\end{figure*}

As noted in \S~\ref{sec:initial_model}, we started with a small
velocity perturbation near the center of the star and the velocity
otherwise zero.  As the simulation begins, reactions heat the
core of the star, and since the background of the star is isentropic,
the heated fluid at the core begins to buoyantly move radially
outward.  Figure~\ref{fig:vort} shows the magnitude of the vorticity
($|\nablab \times \ubold|$) in the three orthogonal slice planes through the
center of the star at several different times.  
At early times, we see
convective flow developing near the center of the star.
By 400~s, the convective flow has grown to fill
the convectively stable region, and we see gravity waves excited
in the stable region above.  The later times show the convective pattern continuing
to strengthen, with small asymmetries in the the vorticity moving
through the inner convective region.  For most of the simulation, we see a
sharp distinction in the character of the flow at the boundary of the
isentropic region in the star.  However, toward the very end of the calculation,
as shown in the very last pane of Figure~\ref{fig:vort}, we no longer
see the separation between the two regions, and the convective plumes appear
to travel through the entire star.

\begin{figure*}
\begin{center}
\plotone{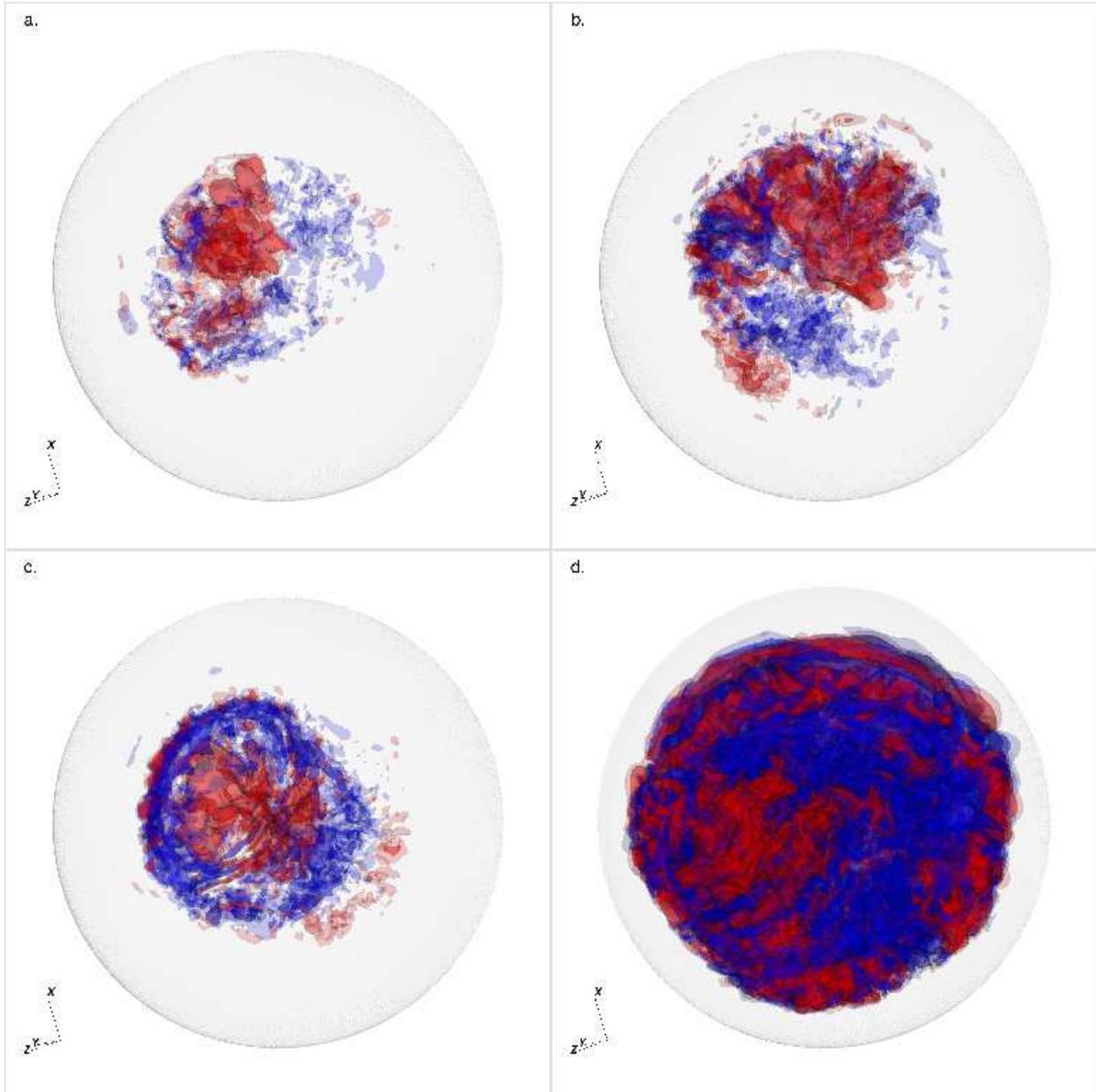}
\end{center}
\caption{\label{fig:radvel} Radial velocity shown at 4 different times:
a.\ 800~s; b.\ 3200~s; c.\ 3420~s; d.\ 7131.79~s.  The latter time corresponds
to the point of ignition.  Red contours indicate outward moving fluid while
blue contours indicate inward moving fluid.  Two contour levels are used
for each sign, $\pm 1.2\times 10^6~\cms$ and $\pm 2\times 10^6~\cms$.  The gray contour is a surface of constant density, $\rho = \rhocutoff$, marking the surface of the star.}
\end{figure*}

Figure~\ref{fig:radvel} shows contours of the radial velocity at
four different times.  Qualitatively, these times represent the early
period (panel a, 800~s), two intermediate snapshots (panels b and c,
3200 and 3420~s respectively), and the very late stage of the
calculation (panel d, 7132~s).  Red indicates fluid moving radially
outward, and blue indicates fluid moving radially inward.  The gray surface
is drawn at a constant density ($\rho = \rhocutoff$) and represents the
surface of the star.   Very early
we see the distinct asymmetric nature to the flow characteristic of a
dipole flow.  The dipole is not nearly as symmetric as that shown in
\citet{kuhlen-ignition:2005}, perhaps due to differing resolution or
the inclusion of the stably-stratified layer surrounding the
convective region in our study.  In general, the outward moving
fluid appears more coherent then the inward moving fluid.  Comparing the images at different
times, we see that the dipole direction changes with time.
Occasionally, the flow takes on a more organized form, with the
inward moving fluid forming a concentric ring around the outward flow,
as shown in panel c.  At the very late stages of the simulation (panel
d), we see what appears to be a breakdown in the distinction between
the stable and unstable regions, with the flow much less
organized and filling most of the volume of the star.  
The narrow gap between the velocity contours and the surface of the star
at late times arises from our sponging term.  We look at the sensitivity
of the results to the position of the sponge in the next subsection. 

\begin{figure*}
\begin{center}
\epsscale{1.0}
\plotone{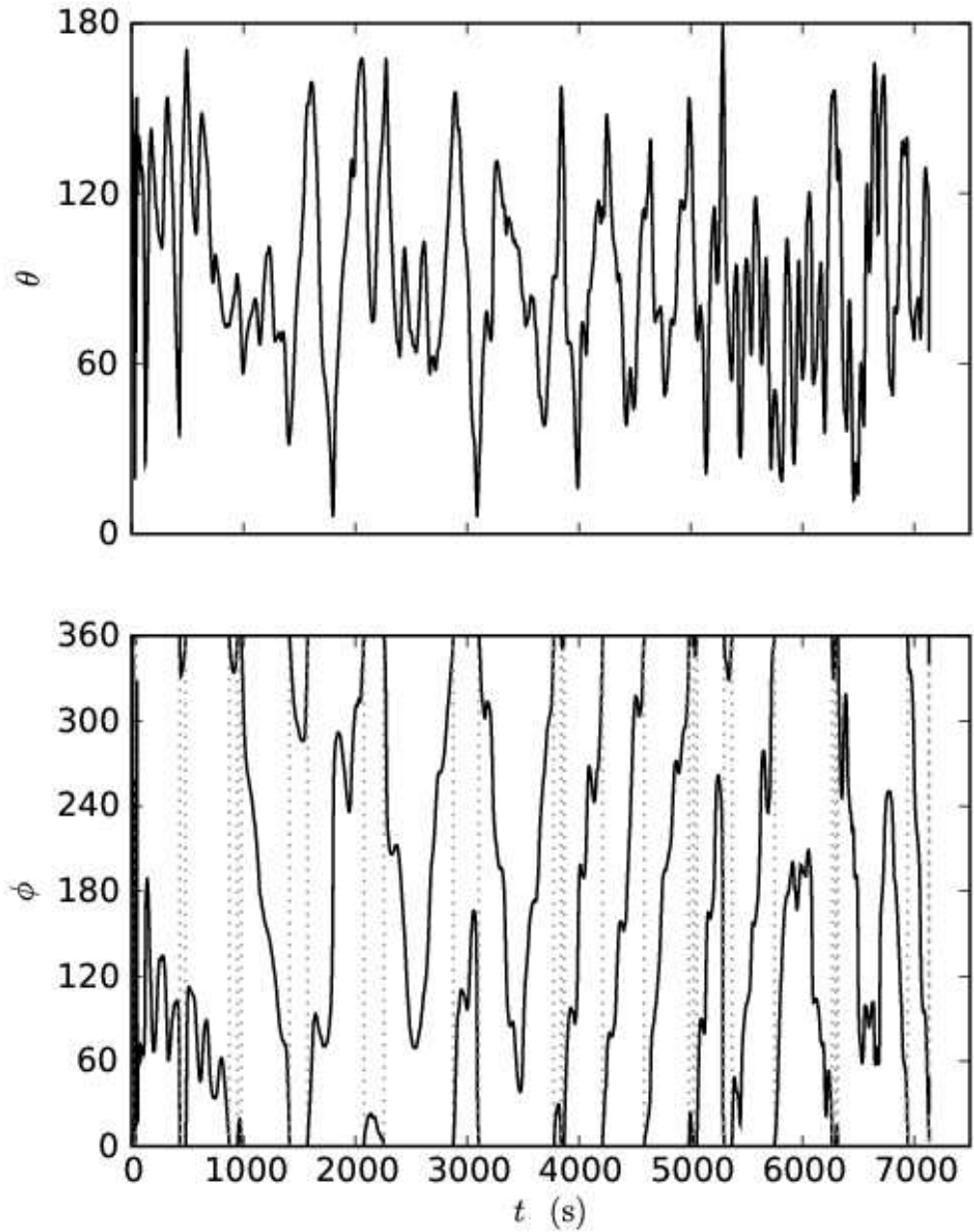}
\epsscale{1.0}
\end{center}
\caption{\label{fig:radvelangles} The spherical angles, $\theta$ and
  $\phi$, computed from $\favgradvel_i$ as a function of time for the
  384$^3$ calculation.  The vertical dotted lines represent where the 
  angle $\phi$ crosses the 0--360$^\circ$ boundary, and are not 
  really discontinuities.  We see that the direction of the dipole changes
  constantly throughout the simulation.}
\end{figure*}

To get a better feel for the change in direction of the dipole, we
compute the spherical angles, $\theta$ and $\phi$, from
$\favgradvel_i$, as defined above.  Figure~\ref{fig:radvelangles} shows these angles as a
function of time.  We see that both angles move through their full
range many times over the course of the simulation.  
We see that the characteristic timescale for $\phi$ to complete a
circuit through $2\pi$ is between 500 and 1000~s.  At late times,
it appears that the dipole is changing direction with a faster period,
especially in the $\theta$ plot.

\begin{figure*}
\begin{center}
\epsscale{1.0}
\plotone{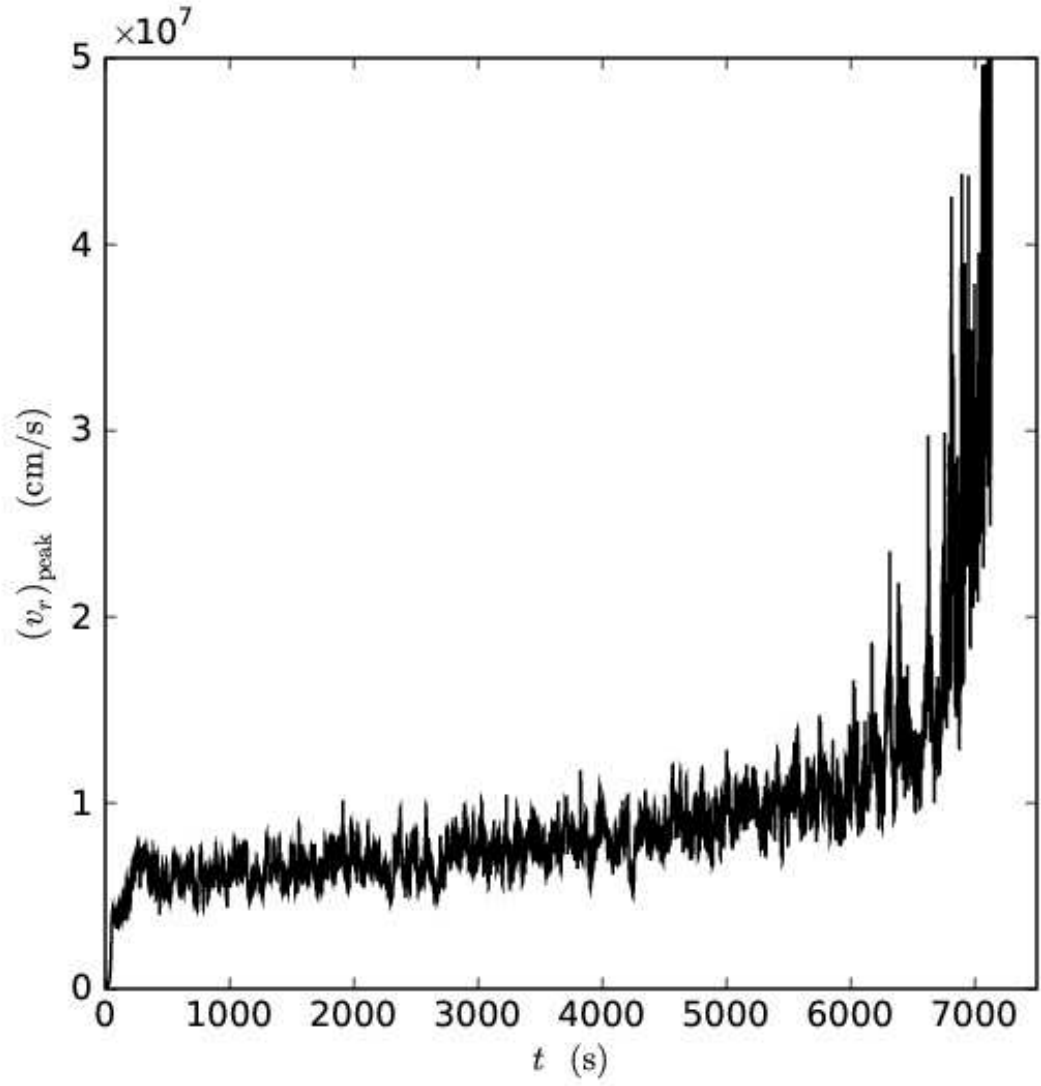}
\epsscale{1.0}
\end{center}
\caption{\label{fig:max_velr} The peak radial velocity as a function
  of time for the 384$^3$ calculation.}
\end{figure*}

Figure~\ref{fig:max_velr} shows the peak radial velocity,
$(v_r)_\mathrm{peak}$, inside the star, as a function of time.  
We see that it slowly rises with time, with a typical peak radial
velocity of $\sim
10^7~\cms$.  Taking the convective region to have a radius of $R_\mathrm{conv} \sim
10^8~\mathrm{cm}$, we define a lower bound to the convective turnover time of
$2 R_\mathrm{conv} / (v_r)_\mathrm{peak} = 20~\mathrm{s}$.

\begin{figure*}
\begin{center}
\epsscale{0.9}
\plotone{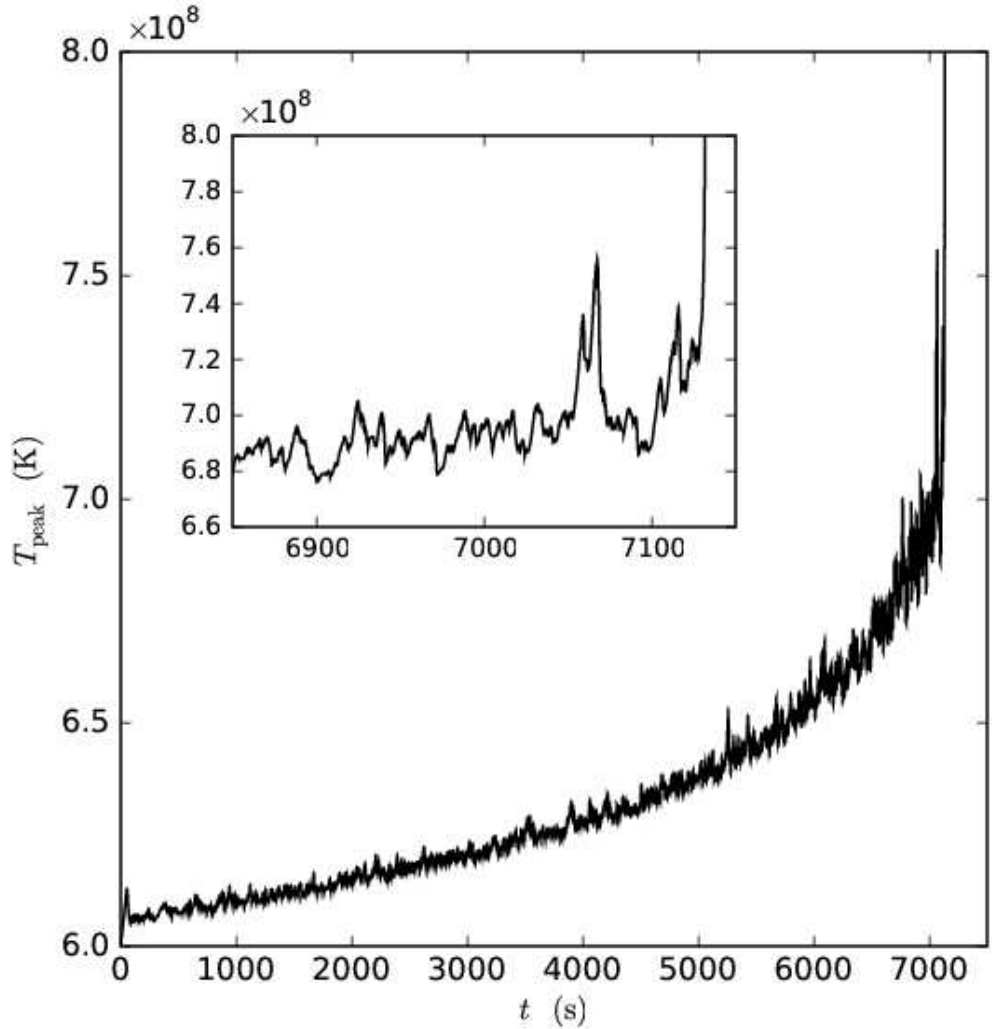}
\epsscale{1.0}
\end{center}
\caption{\label{fig:max_temp} The maximum temperature in the white dwarf
as a function of time for the 384$^3$ calculation.  The temperature
increase is highly nonlinear, ending at ignition.  
To show detail, we restrict the vertical range of the plot to $8\times 10^{8}$~K.
The inset shows the structure of $T_\mathrm{peak}$ during the last $\sim$200~s.
We see large, but damped excursions in central temperature just prior to ignition.
}
\end{figure*}

Ignition will occur when the reactions proceed so strongly that hot,
reacting bubbles are not quenched by adiabatic expansion in the
convective motions carrying the
fluid away from the center of the star.  Since the $^{12}\mathrm{C} + \, ^{12}\mathrm{C}$
reaction rate is so strongly temperature-sensitive, the peak
temperature in the star serves as a good guide for observing the 
progression toward ignition.  Figure~\ref{fig:max_temp} shows the
peak temperature as a function of time for this calculation.  We see
a short transient at the start of the calculation where the temperature
quickly rises and then settles back down---this occurs from the nonlinear
feedback of the temperature into the reactions when the flow field
is not yet fully developed.  After a short amount of time, a convective
flow field develops that properly matches the energy generation at 
the center of the star, and the temperature settles into
a long, gradual rise.  About halfway through the calculation, we can
clearly see that the temperature rise is non-linear, and the temperature
increase accelerates toward the very end, up to the point of ignition.
The inset in Figure~\ref{fig:max_temp} shows the behavior of $T_\mathrm{peak}$
during the last 200~s.

\begin{figure*}
\begin{center}
\epsscale{0.9}
\plotone{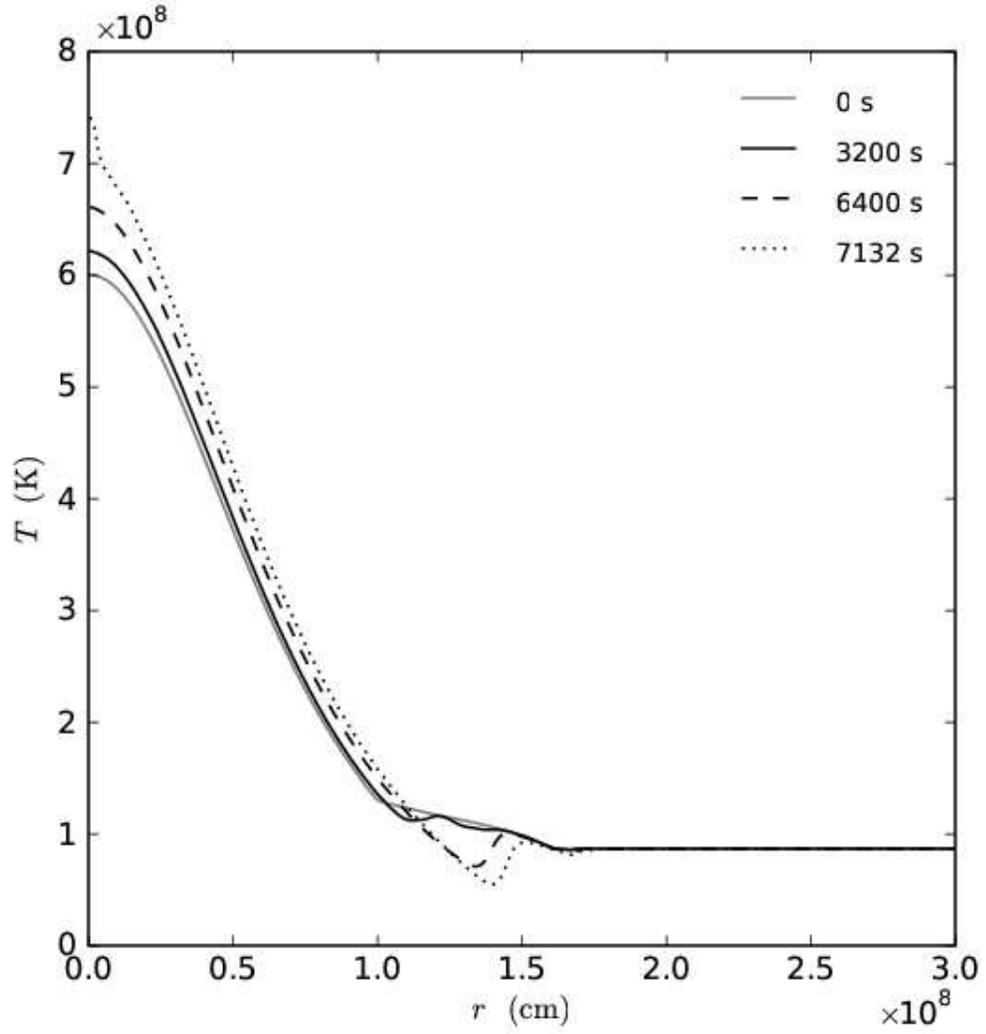}
\epsscale{1.0}
\end{center}
\caption{\label{fig:temp_averages} Average temperature as a function of radius
in the white dwarf, shown at the initial time and 3 later times, for the 384$^3$ 
convection calculation.  With time, the energy dumped into the star by 
reactions causes the temperature to increase throughout the star.  The curve at
7132~s corresponds to the time of ignition.}
\end{figure*}

\begin{figure*}
\begin{center}
\epsscale{0.9}
\plotone{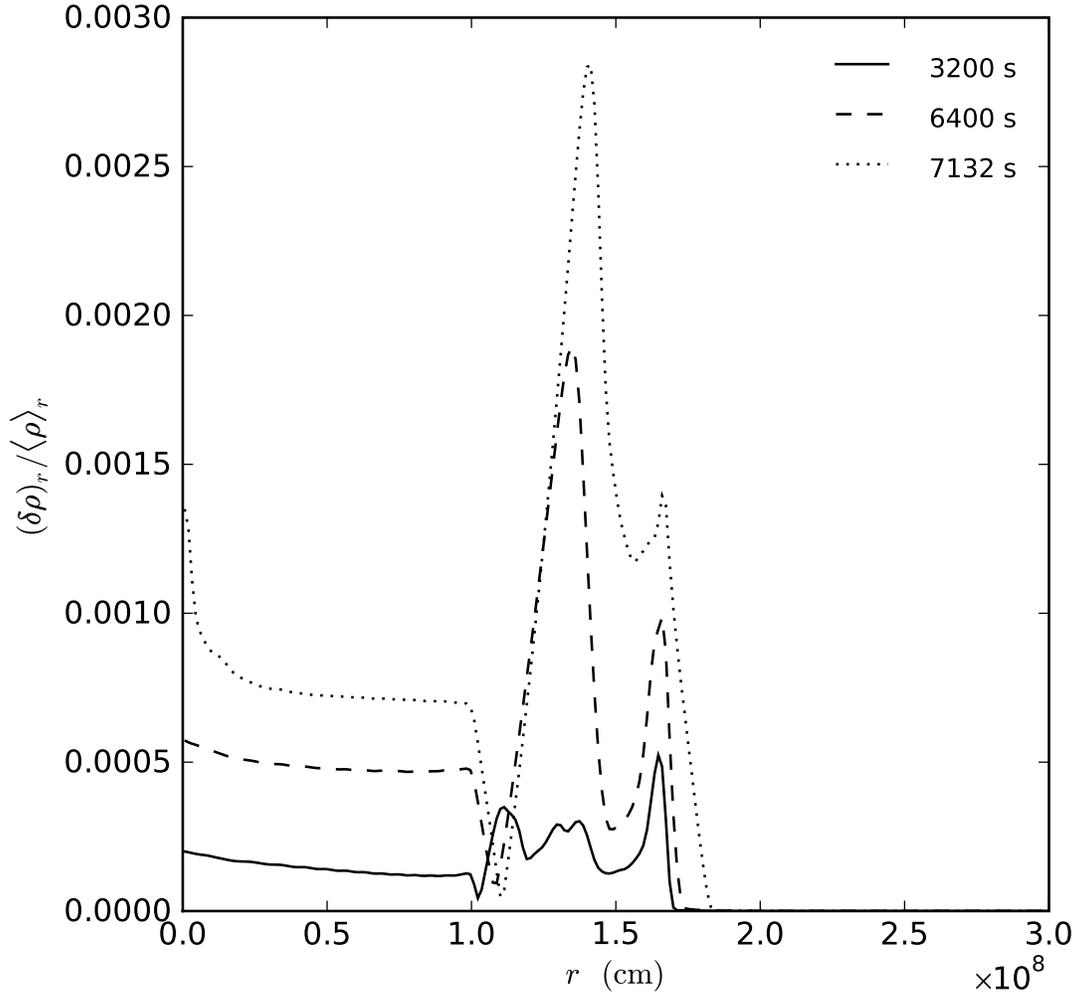}
\epsscale{1.0}
\end{center}
\caption{\label{fig:deltarho} $(\delta \rho)_r / \langle \rho
  \rangle_r$ vs.\ $r$ for the 384$^3$ white dwarf convection problem at 3 times.  The
  curve at 7132~s corresponds to the time of ignition.  We see that at all times, 
  $(\delta \rho)_r / \langle \rho \rangle_r$ remains well below 1\% everywhere inside the
  star.}
\end{figure*}

The reactions dump energy into the star, and it heats up throughout.
Figure~\ref{fig:temp_averages} shows the average temperature at a given
radius as a function of radius at several different times.  As we see,
the temperature increases throughout the convective region.  At late
times, we see a distinct change in the temperature structure at the
boundary of the convective region.  This change in temperature
structure corresponds to the penetration of the vortical flow through
the original boundary between the stable and unstably-stratified
regions in the vortical plot (Figure~\ref{fig:vort}) shown above.  It
is not clear how robust this change in the character of the flow is to
resolution---that is something that will be explored through higher
resolution studies in the future.  It is also the case that those
outer layers, near the transition to a stably-stratified fluid, are
where we would expect the expansion of the star to be greatest, so we
need to check if neglecting the base state evolution was warranted.
Figure~\ref{fig:deltarho} shows $(\delta \rho)_r / \langle \rho
\rangle_r$ vs.\ radius at several times.  This is a measure of how
much expansion of the star has taken place.  If $(\delta \rho)_r$ is
large compared to $\langle \rho \rangle_r$, then the full state is
carrying the expansion rather than the base state, and as we've shown
in paper~II, this can lead to inaccuracies.  As
Figure~\ref{fig:deltarho} shows, $(\delta \rho)_r / \langle \rho
\rangle_r$ is always below 1\%, indicating the departure from the base
state is small, and any expansion would be minimal.  
In each case, the curve at 7132~s corresponds to the point is ignition, 
discussed below.

We can also look at the total kinetic energy in the star, which we 
compute as
\begin{equation}
K = \sum_{\Omega_\mathrm{star}} \rho |\ubold|^2 \Delta x \Delta y \Delta z \enskip .
\end{equation}
At the point when the peak temperature reaches $8\times
10^8$~K, the total kinetic energy inside the star is $6.24\times
10^{46}~\mathrm{erg}$.  To put this in context, we can compare the
gravitational potential energy of the star, defined from our base
state as
\begin{equation}
U = -\int_{\Omega_\mathrm{star}} \frac{G M(r) dM}{r} \enskip ,
\end{equation}
with $dM = 4\pi r^2 \rho_0 dr$.  For our model, the gravitational
potential energy is $-3.2 \times 10^{51}~\mathrm{erg}$.  The
internal energy of the gas is also quite large, 
$E_\mathrm{int} = 2.7\times 10^{51}~\mathrm{erg}$, giving an
energy difference of $\sim -5\times 10^{50}~\mathrm{erg}$ that
needs to be overcome to unbind the star.  
Therefore,
kinetic energy release up to the point of ignition is a tiny fraction
of what is needed to unbind the star---as expected. 

The $384^3$ calculation took 113156 time steps to reach a simulation
time of 7131.8~s---at which point the peak temperature had risen
to $8\times 10^8$~K, and ignition shortly followed.  Overall, the
average time step is 0.063~s.  At this same instant, the Mach number,
attained in the outer layers of the star, reached a value of 0.079.
Earlier in the calculation, the maximum Mach number in the domain
was considerably lower.  For comparison, the highest sound speed in
the star (at its center) is $9.5\times 10^8~\cms$, which would give
a corresponding time step of $7\times 10^{-4}$~s (assuming a
CFL number of 0.5, and $|U| \ll c_s$, where $c_s$ is the sound speed).

\subsubsection{Effect of $\rhocutoff$}

As we noted above, the behavior of the coefficient in our constraint
term and the location of the sponge are set by the density we refer to
as $\rhocutoff$.  To assess the influence of our choice of
$\rhocutoff$, we perform a pair of simulations on a $256^3$ grid that
are identical except for the value of $\rhocutoff$.  For our control
case, we use $\rhocutoff = 3\times 10^6~\gcc$, the value chosen for
our main calculation.  To explore the effects of lowering $\rhocutoff$,
we also try a value of $\rhocutoff = 10^6~\gcc$.  We note that the
location of the sponge in the momentum equation remains keyed to the
choice of $\rhocutoff$, so with the lower value of $\rhocutoff$, the
location of the start of the sponge moves outward from the center of
the star.  In terms of mass, $\rhocutoff = 10^6~\gcc$ means that the
mass of the star enclosed is $1.381~M_\odot$, compared to
$1.378~M_\odot$ with $\rhocutoff = 3\times 10^6~\gcc$.  The location
of the start of the sponge contains $1.363~M_\odot$, compared with
$1.320~M_\odot$ with $\rhocutoff = 3\times 10^6~\gcc$.

\begin{figure*}
\begin{center}
\epsscale{1.0}
\plotone{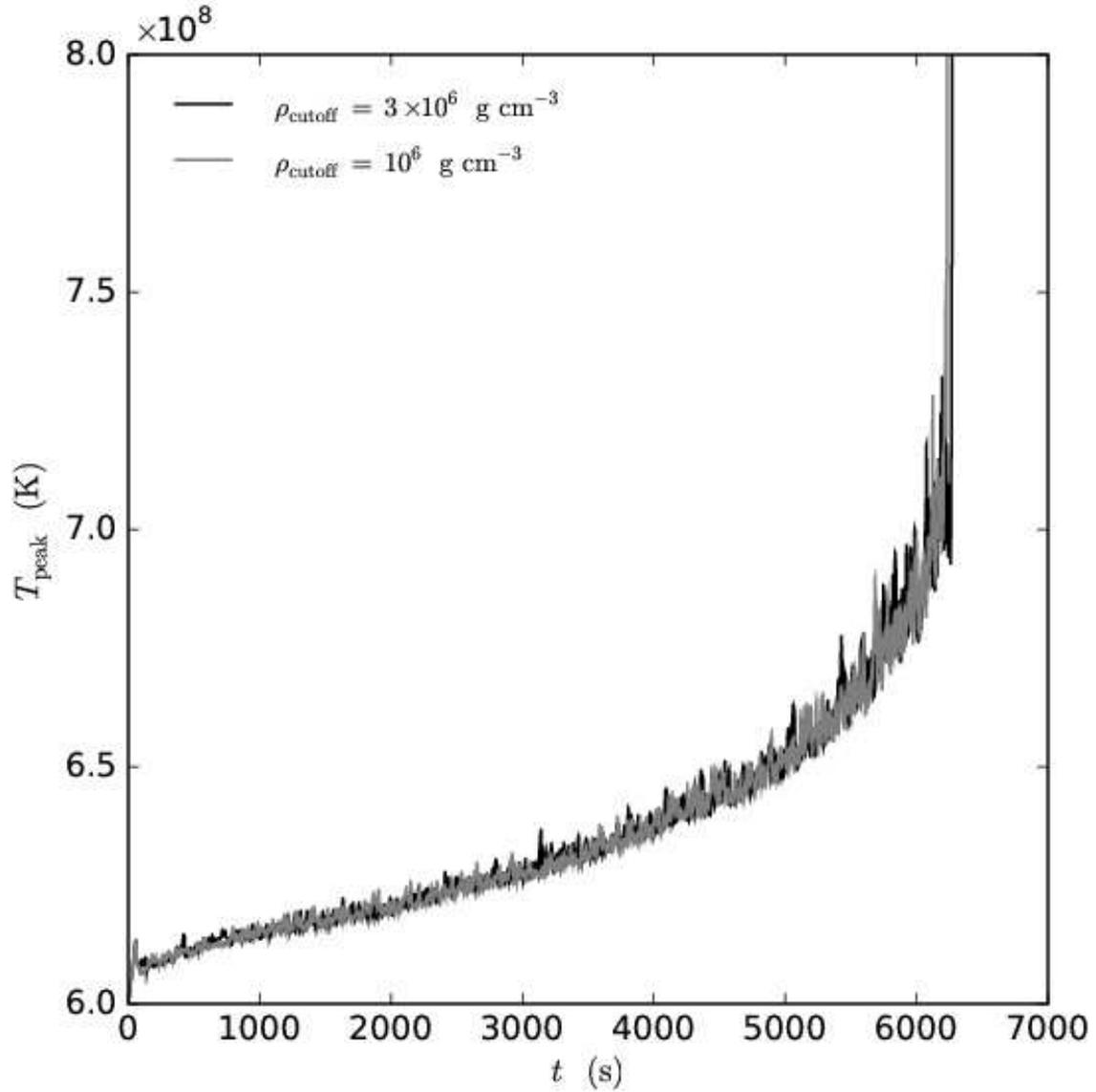}
\epsscale{1.0}
\end{center}
\caption{\label{fig:anelastic_cutoff} The maximum temperature in the white dwarf
as a function of time for the two different choices of $\rho_\mathrm{cutoff}$ 
($10^6~\gcc$ and $3\times 10^6~\gcc$).  Both
simulations use a $256^3$ grid.  Here we see excellent agreement between
the two cases, indicating that the peak temperature is insensitive to
our choice of $\rhocutoff$.}
\end{figure*}

Figure~\ref{fig:anelastic_cutoff} shows $T_\mathrm{peak}$ as a function
of time for the two calculations.  As we see, the two curves track very
well, indicating that the choice of $\rhocutoff$ has little influence
on the temperature behavior near the center of the star.  The time at
which final ignition occurs differs between these two cases by only
38.4~s out of over 6000~s of evolution.

As noted above, the choice of $\rhocutoff$ is used to prevent the
velocities from growing too large as the fluid experiences the steep
density gradient at the edge of the star.  We note that the time step
the code takes with $\rhocutoff = 3\times 10^6~\gcc$ is 23\% larger
then with $\rhocutoff = 10^6~\gcc$.  Thus it is computational
favorable to use the slightly higher value of the cutoff density.

\subsubsection{Ignition}

\begin{figure*}
\begin{center}
\epsscale{0.8}
\plotone{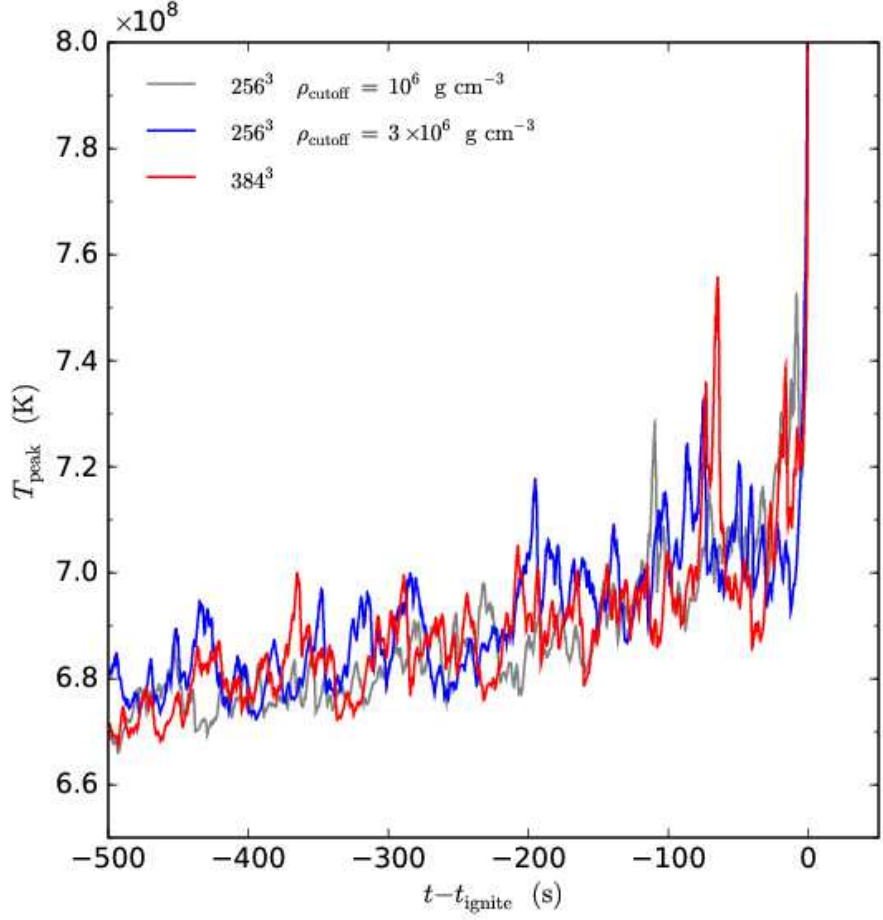}
\epsscale{1.0}
\end{center}
\caption{\label{fig:Tcompare} The peak temperature, $T_\mathrm{peak}$, as a function of time
for the high resolution run ($384^3$) and two medium-resolution runs ($256^3$), offset so
the time of ignition lines up.  Here we show only the last 500 s leading up to ignition.  We see that the temperature rise approaching ignition matches
well between these different calculations.}
\end{figure*}

As the inset in Figure~\ref{fig:max_temp} shows, up to the point of
ignition, the peak temperature rises rapidly, only to fall again,
as a spark fails to ignite.  Figure~\ref{fig:Tcompare} shows the
temperature structure in the last 500~s for the $384^3$ calculation
and both $256^3$ calculations, shifted so the time of ignition lines up.
All three runs show the peak temperature fluctuating rapidly before
ignition, indicating some hot spots failed to ignite. 
In each case, eventually, a hot spot burns faster than it
cools and the temperature rapidly shoots up to over
$10^{10}~\mathrm{K}$---igniting the first flame.  At this point, our
algorithm cannot deal with the rapid energy release, and the low Mach
number approximation breaks down, so we stop the calculation.  Physically,
at this point the nuclear burning timescale is much shorter than the 
advection timescale.  Whether
a second hot spot ignites shortly following this one is not something
we can address presently.  

At the point of ignition, the radial velocity
(Figure~\ref{fig:max_velr}) rises rapidly, reaching
unphysically high values post-ignition---a symptom of the breakdown
of the low Mach number method when the first flame ignites.  However,
prior to ignition, the algorithm remains valid, and we see that the
radial velocities rise to $\sim$ few $\times 10^7~\cms$.  These
velocities will affect the dynamics of the first flames to ignite.

To determine the location of the ignition, we compute the radius of
the hot spot that ignited by looking at the peak temperature.  We
define the ignition radius as the position of the hot spot at the time
when the peak temperature passes beyond $8\times 10^8~\mathrm{K}$.
In all cases, once this temperature is exceeded, the temperature quickly
shoots up to $O(10^{10})$~K.

For the high-resolution calculation, we find the location of ignition
to be 21.6~km from the
center of the star.  The location of the peak temperature remained
steady from (and possibly before) about 0.8~s before the peak
temperature rose above $8\times 10^8$~K, indicating that the hot spot
was not moving very fast.  The radial velocity in the zone with the
peak temperature at the time we satisfy our ignition criteria is
only 4.8~km/s. 
We note that physical center of the star is
on a vertex on our Cartesian grid, so central ignition in this
simulation would be at $\sqrt{3} \Delta x / 2 = 11.3~\mathrm{km},$
the distance from the center of the closest grid cell to the center
of the star.

\begin{figure*}
\begin{center}
\epsscale{1.0}
\plotone{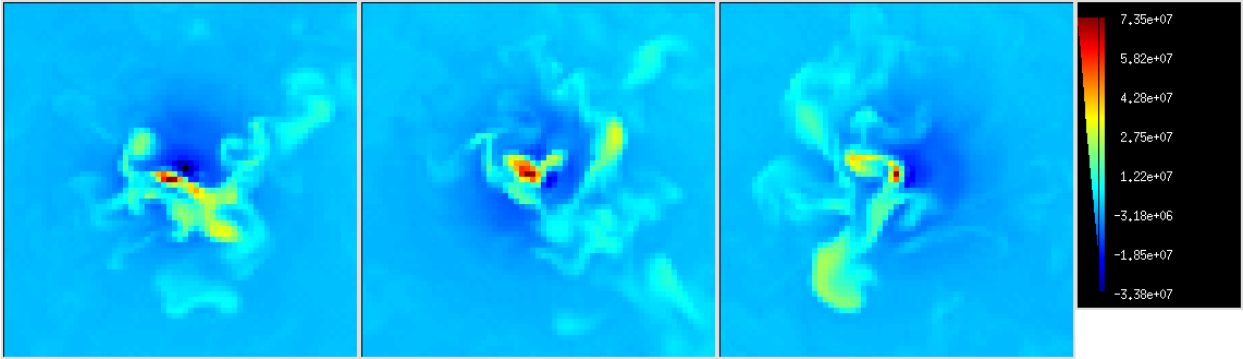}
\epsscale{1.0}
\end{center}
\caption{\label{fig:tpert} The perturbational temperature ($T -
  \langle T \rangle$) in three orthogonal slice planes ($x$-$y$,
  $x$-$z$, and $y$-$z$) passing through the point $(2.48\times
  10^8~\mathrm{cm},\, 2.49\times 10^8~\mathrm{cm},
  \, 2.52\times 10^8~\mathrm{cm})$, at a simulation
  time of 7131.79~s.  Only the central $64^3$ portion of the 
  domain is shown (833.3 km on a side).  Ignition occurs in the zone
  where $T - \langle T \rangle$ is the largest. }
\end{figure*}

Figure~\ref{fig:tpert} shows the perturbational temperature (full state
temperature, $T$, minus the average temperature at the corresponding
radius, $\langle T \rangle$) in the central $64^3$ zones (833~km on a
side).  We've picked the location of the slice planes to cut right
through the hot spot.  As the figure shows, the peak temperature is
strongly localized to a single zone, with an extended hot region
surrounding this location.

For the two medium resolution ($256^3$) cases, we find the location of
the hot spot when $T_\mathrm{peak}$ crosses $8\times 10^8$~K to be
84.5~km ($\rhocutoff = 3\times 10^6~\gcc$) and 32.4~km ($\rhocutoff =
10^6~\gcc$).  For reference, central ignition at this resolution would
correspond to 16.9~km.  For the $\rhocutoff = 3\times 10^6~\gcc$ case,
the location of the peak temperature changes rapidly (moving outward
from the center) as the peak temperature crosses $8\times 10^8$~K,
ranging from 32.4~km when $T_\mathrm{peak} = 7.51\times 10^8$~K at
6267.6~s to 89.0~km when $T_\mathrm{peak} = 8.13\times 10^8$~K at
6269.9~s.  The radial velocity in the zone satisfying our ignition
criteria is 39~km/s.  Clearly, the flow dynamics at the location of
ignition are significant, in this case.  We do not see movement of
this magnitude for the $\rhocutoff = 10^6~\gcc$ case, where the radial
velocity in the zone satisfying our ignition criteria is only
2.9~km/s.  For future calculations we will store the location of the
hot spot along with the peak temperature at every time step to help
better understand these dynamics.

Taken together, we see a distribution of ignition radii in our
results, ranging from 21.6~km to 89.0~km.  Ignition is a highly
nonlinear process, and we expect that were we to perform more runs,
slightly tweaking the initial conditions, we would observe different values,
all of which sample the distribution function of possible ignition
locations in the problem.  Owing to the stochastic nature of the
problem, to really understand the ignition process requires performing
a large number of slightly different calculations to map out the
distribution function.

\subsubsection{Effect of Resolution}

It has been suggested that the behavior of convective flow can
dramatically change in character in flows of high Rayleigh number
\citep{kadanoff}.  In our simulation code, we do not explicitly add
viscosity to the momentum equation (eq.~[\ref{eq:utildeupd}]), so our
Rayleigh number is determined by the numerical viscosity inherent in
our advection scheme.  Furthermore, the nature of the turbulence will
depend on the Reynolds number of the flow, which again in our
simulations is determined by numerical viscosity.  Practically
speaking, the way to increase the effective Reynolds and Rayleigh
numbers of the simulation is to move to higher-order advection
methods and to increase the resolution.

While no amount of resolution will bring our effective Reynolds and
Rayleigh numbers up to the $O(10^{14})$ and $O(10^{25})$ values,
respectively, we expect in the true convecting white dwarf
\citep{Woosley:2004}, it is interesting to look at how the general
results change with resolution.  A second reason to explore resolution
is that it is not known what size region will ignite.  One might
imagine that a region the size of only a few flame thicknesses needs
to heat up to ignite a flame.  At the central densities in the white
dwarf, the flame thickness is $O(10^{-4}~\mathrm{cm})$
\citep{timmeswoosley1992}---this is far below any resolution that can
be obtained by a large-scale simulation code.  However, the flame will 
initially burn in place, growing until it is large enough
(about 1~km) that buoyancy becomes significant and it begins to rise
and deform \citep{zingaledursi}.  While this is still a smaller length
scale than considered here, it is not out of reach with mesh refinement
and larger computers.

To begin to understand the effect of resolution, we consider three
cases: $128^3$, $256^3$, and $384^3$, corresponding to physical zone
sizes of 39.1, 19.5, and 13.0~km respectively.  We note that for the
present study, computer resources prevent us from considering a
$512^3$ or higher case.

\begin{figure*}
\begin{center}
\epsscale{0.8}
\plotone{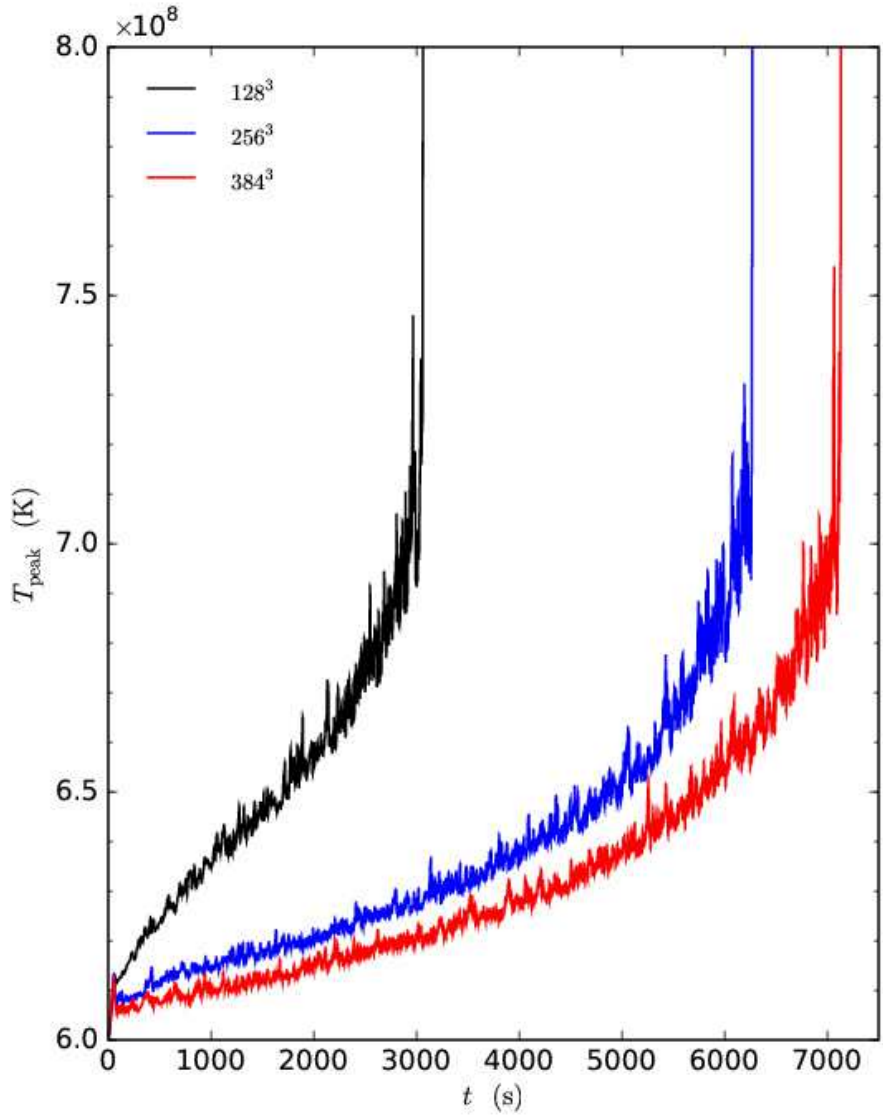}
\epsscale{1.0}
\end{center}
\caption{\label{fig:resolution} The peak temperature, $T_\mathrm{peak}$, in the white dwarf
as a function of time for three different resolutions.  We see that as we
increase the resolution, the temperature increase is slower.  The lowest
resolution case reaches ignition very quickly.  Once it ignites, the peak temperature
climbs to $\sim 10^{10}$~K almost instantly.  
To show detail, we restrict the vertical
range of the plot to $8\times 10^{8}$~K.}
\end{figure*}

\begin{figure*}
\begin{center}
\epsscale{0.8}
\plotone{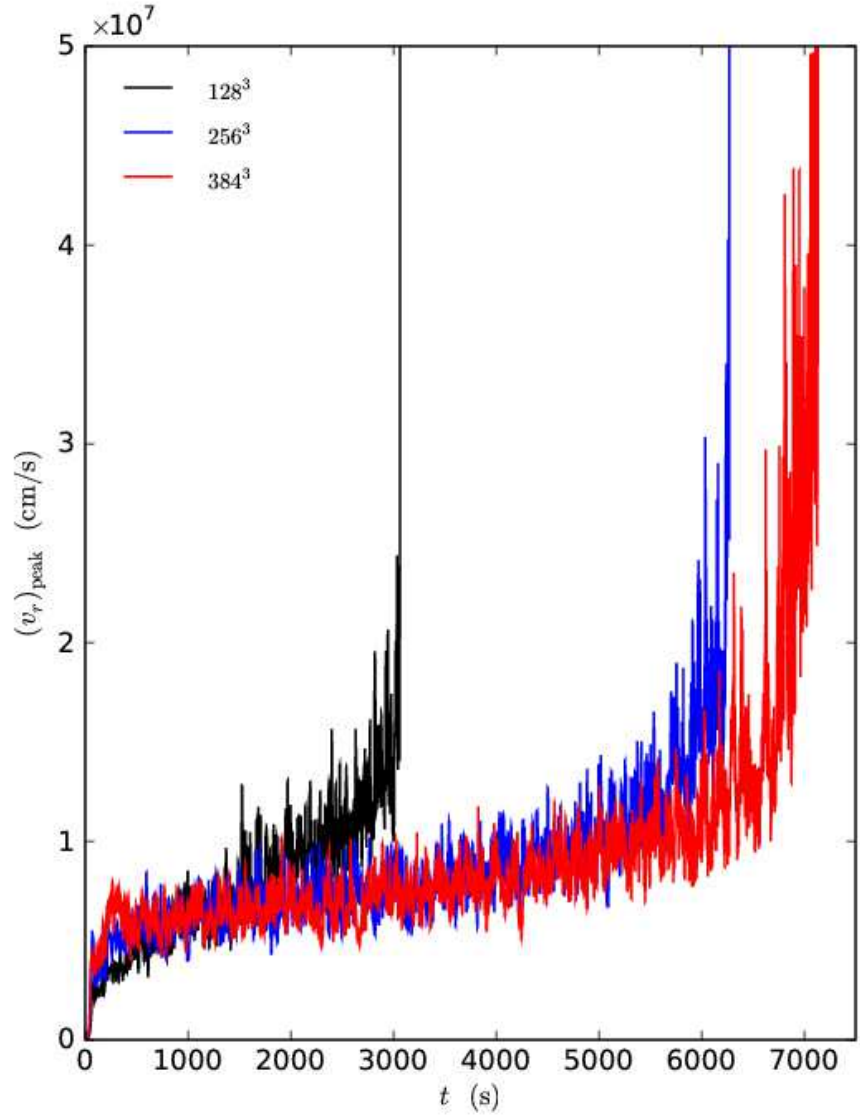}
\epsscale{1.0}
\end{center}
\caption{\label{fig:resolution_velr} The peak radial velocity, $(v_r)_\mathrm{peak}$,
in the white dwarf
as a function of time for three different resolutions.  }
\end{figure*}

Figure~\ref{fig:resolution} shows $T_\mathrm{peak}$ vs.\ time for the
three different resolutions.  Immediately we see that the $128^3$ case
reaches ignition much faster than the two higher resolution cases.  In fact,
from our starting temperature of $6\times 10^8$~K, the highest resolution
run takes more than twice as long in simulation time to reach ignition.
Also apparent in the coarsest resolution run is that the temperature
did not drop after the initial transient---this contributes to the
faster overall evolution.  Both of the higher resolution cases see a
drop in the temperature after the initial transient, as the developing
velocity field carries the heated fluid away from the center of the star.
Close to ignition, the $256^3$ and $384^3$ runs show a similar slope
in the $T_\mathrm{peak}$ vs.\ $t$ curve shown in Figure~\ref{fig:Tcompare}. 
The peak radial velocity as a function of time also shows differences
between the resolutions, as shown in Figure~\ref{fig:resolution_velr}.
There does appear to be some convergence with resolution.

\section{Conclusions and Discussion}

We have demonstrated that our simulation code, MAESTRO, is capable of
following the convective flow in a white dwarf leading up to the
ignition of a Type Ia supernova.  We have explored the 
sensitivity of the results to resolution and to the choice of 
low density cutoff, 
$\rhocutoff$.  Our test problem shows that discretizing
the star on a Cartesian grid with a radial base state leads to
an accurate representation of the flow.

Over many convective turnover times, our simulations capture the rise
of the peak temperature in the white dwarf up to ignition, recover the
dipole nature of the convective flow first shown in
\citet{kuhlen-ignition:2005}, and track the change in direction of the
dipole.  We see, for the first time in multidimensional simulations,
the distinct change in the nature of the flow at the outer boundary of
the convective region, as discussed in \citet{pirochang}.  The late time 
breakdown of this interface needs further investigation. 

All of our models reached ignition.  For the two medium resolution runs,
the ignition occurred at a radius of 32.4 and 84.5~km.  For the high resolution
run, it occurred at 21.6~km.  These are the locations of the first flames.  
We note that this is a highly-nonlinear problem, 
and small changes in the state of the star could affect the ignition
process.  
To really understand the statistical distribution of initial
ignition points requires running a suite of calculations, varying the
initial model (central density, size of initial convective region),
and the initial state of the star.  With an ensemble of such calculations,
we could get a much better understanding of the ignition process.
We also need to understand how the ignition process differs with 
higher resolution. 

A detailed comparison to \citet{hoflichstein:2002} or
\citet{kuhlen-ignition:2005} is difficult, because of the differing
geometries used.  \citet{hoflichstein:2002} (hereafter HS) simulated a
90$^\circ$ wedge in 2-d, but cut out the innermost 13.7~km (in their
``extended computational domain'' run).  As we discussed, in the
present calculation the burning is strongly peaked near the center of
the star, so cutting out the center would miss a great deal of the
energy generation.  It would also prevent the fluid from flowing
through the center of the star, which is the dominant pattern seen in
the present calculation.  Because HS used a spherical grid, and
allowed the radial spacing to vary, there is no single grid resolution
for their simulation, but they state that near the inner boundary, the
grid resolution is ``$\sim 2$ km''---about $6\times$ finer than the
uniform resolution we use throughout the star.  In both their
calculation and our calculation, several hours of the runaway are
followed leading up to the point of ignition ($\sim 3$~hours for HS,
$\sim 2$~hours for our calculation).  Also, in both cases, the
ignition takes place in a single zone.  However, the details of the
ignition differ---because HS have an inner boundary and a wedge-shaped
domain, compression is generated that leads to the ultimate ignition
near the center.  In our calculation, we have flow through the center throughout the
simulation.  HS found the ignition to take place at a radius of
27~km in their model, which is within the range we report in our study.
Furthermore, because their initial model had a strong gradient in the
carbon mass fraction, with the outer portion of the star having a
carbon mass fraction of 0.4 and the core having a value close to
0.25 (see HS, figure 1), HS report that the expanding convective region results
in an increase in 
the carbon mass fraction at the center from 0.25 to 0.36.  This is not
the case in our model, where the carbon burning caused a 0.5\%
decrease in the central carbon mass fraction over the time we modeled.
HS quote convective velocities between 40 and
120~km~s$^{-1}$ at ignition.  In our case, the radial velocities were
around 100~km~s$^{-1}$ for most of the evolution, rising to several
times that just before ignition.

The calculation by \citet{kuhlen-ignition:2005} (hereafter KWG) modeled 
the star in 3-d, with an inner boundary at 50~km, again cutting out the
center of the star.  They also put the outer boundary at 500~km, which
is approximately halfway through the convectively unstable region in
the initial model (KWG and the present study use very similar
initial models,
generated from the Kepler stellar evolution code).  Despite cutting
out the center, KWG found that a large-scale dipole flow pattern
dominates---similar to what we see (although we note that they also
did a rotating model, and saw the dipole break down).  Unlike HS or
the present calculation, KWG do not model the evolution continuously
leading up to ignition, but rather model two snapshots in time,
corresponding to central white dwarf temperatures of $7\times 10^8$~K
and $7.5\times 10^8$~K, with durations of 70 and 41~s respectively.
This amounts to a few turnover times \citep{kuhlen-ignition:2005}.  It
is difficult to compare resolution with KWG, as they use a spectral
method for the spatial discretization.  KWG quote typical velocities
of 50 - 100~km~s$^{-1}$---consistent with both HS and the present
calculation.  Finally, in contrast to both HS and our calculation,
KWG did not follow the evolution to the ignition of the first flame,
but rather inferred from the size of the dipole flow pattern that
ignition would likely be off-center.  
Overall the comparison to these previous calculations, and the variation
seen in our own set of calculations, suggest that more full-star, 
three-dimensional calculations, with varying initial parameters, 
are needed, to fully understand the ignition process.

Future work will include both algorithmic improvements and more
realistic physics.  The focus of the next set of calculations will be
to continue exploring the nature of the convection.  On the
algorithmic front, we will switch to an unsplit implementation of the
piecewise parabolic method \citep{ppm,ppmunsplit} for the advection
scheme and begin to incorporate adaptive mesh refinement.  Together
with an increase in grid resolution, these changes will allow us to
push to higher effective Reynolds numbers.  Physically, we will
improve the nuclear energetics, add an enthalpy equation to better
define the temperature throughout the simulation, incorporate the
expansion of the base state, and include rotation.  Numerical
\citep{kuhlen-ignition:2005} and analytic \citep{piro2008} work has
shown that the effects of rotation can be significant.

The present calculations show only the ignition of the first flame,
but the convection continues and it is likely that other flames
will ignite.  This process of ongoing ignition could be critical
to the understanding of SNe Ia.
To date, only limited studies \citep{schmidtniemeyer2006} have been
performed investigating the effects of temporally-spaced ignition
spots.  By capturing the flames that ignite in our simulations
and propagating them in a controlled fashion, 
we can continue a convective calculation 
to simulate the formation of additional ignition spots.

Presently, our algorithm is unable to follow the evolution past the
point where ignition occurs.  However, right up until that point, the
model remains valid.  Therefore, these models can still provide useful
starting conditions for explosion models run with fully compressible
codes, simply by mapping the fully convective state right before
ignition into a 3-d compressible code.  On the longer term, an
extension of our method to include long wavelength acoustics would
extend the validity of this method to $M \sim 1$ (see, for example,
\citealt{gattibonocolella}).
We will also work on incorporating a flame model to capture 
the ignition of the flame(s) on the grid, so that we may
continue the convective calculation in the presence of these first
flames.

\acknowledgments

We thank Frank Timmes for making his equation of state routines
publicly available.  We thank Alan Calder, Jonathan Dursi, and Chris
Malone for many useful discussions on this work.  Finally, we thank
Mike Lijewski for his continued help on software development.  The
work at Stony Brook was supported by a DOE/Office of Nuclear Physics
Outstanding Junior Investigator award, grant No.\ DE-FG02-06ER41448,
to Stony Brook.  The work at LBNL was supported by the SciDAC Program
of the DOE Office of Mathematics, Information, and Computational
Sciences under the U.S. Department of Energy under contract
No.\ DE-AC02-05CH11231.  The work at UCSC was supported by the DOE SciDAC
program, under grant No.\ DE-FC02-06ER41438.

Computer time for the main calculation in this paper was provided
through a DOE INCITE award at the Oak Ridge Leadership Computational
Facility (OLCF) at Oak Ridge National Laboratory, which is supported by the
Office of Science of the U.S. Department of Energy under Contract
No.\ DE-AC05-00OR22725.  We thank Bronson Messer for his help with this
machine.  Computer time for the supporting calculations
presented here was provided by Livermore Computing's Atlas machine
through LLNL's Multiprogrammatic \& Institutional Computing Program.
The test problem calculation used resources of the Argonne Leadership
Computing Facility at Argonne National Laboratory, which is supported
by the Office of Science of the U.S. Department of Energy under
contract DE-AC02-06CH11357.  Some visualizations were performed using
the VisIt package.  We thank Gunther Weber for his assistance with VisIt.

\clearpage


\clearpage

\clearpage

\clearpage

\clearpage

\clearpage

\clearpage

\clearpage

%

\clearpage

\clearpage

\clearpage

\clearpage

\clearpage

\clearpage

\clearpage

\clearpage

\clearpage

\clearpage

\end{document}